\documentclass[twocolumn]{aastex631}

\usepackage{graphicx}
\usepackage{dcolumn}
\usepackage{bm}
\usepackage{subfigure}
\usepackage[nointegrals]{wasysym}
\usepackage{verbatim}
\usepackage{color}
\usepackage{amsmath}
\usepackage[utf8]{inputenc}

\newcommand{\snr}{\ensuremath{S/N}}

\newcommand{\dx}[1]{\mathrm{d}{#1}\,}
\newcommand{\Cltot}[1]{C_{#1}^{\mathrm{tot}}}
\newcommand{\Clfg}[1]{C_{#1}^{\mathrm{fg}}}

\newcommand{\Clhatphi}[1]{C_{#1}^{\hat{\phi}\hat{\phi}}}
\newcommand{\CLhatphiX}[1]{C_{L}^{\hat{\phi},#1}}
\newcommand{\Clphi}[1]{C_{#1}^{\phi\phi}}

\newcommand{\Clhatkappa}[1]{C_{#1}^{\hat{\kappa}\hat{\kappa}}}

\newcommand{\Clkappa}[1]{C_{#1}^{\kappa\kappa}}

\newcommand{\vecl}{\mathbf{l}}

\newcommand{\vecL}{\mathbf{L}}
\newcommand{\Tfg}{T^{\mathrm{fg}}}
\newcommand{\Tcmb}{T_{\mathrm{CMB}}}

\newcommand{\Aphi}{A^{\phi}}

\newcommand{\Alens}{A_{\mathrm{lens}}}

\newcommand{\Tcib}{T_{\mathrm{CIB}}}

\newcommand{\Lmax}{L_{\mathrm{max}}}

\newcommand{\websky}{\textsc{websky}}
\newcommand{\sehgal}{S10}
\newcommand{\nemo}{\textsc{nemo}}

\newcommand\eqn[1]{equation~\ref{#1}}

\renewcommand\fig[1]{Figure~\ref{#1}}

\newcommand\sect[1]{Section~\ref{#1}}
\newcommand\tab[1]{Table~\ref{#1}}
\newcommand\app[1]{Appendix~\ref{#1}}

\newcommand{\planck}{{\it Planck}}

\usepackage{aas_macros}

\begin{document}
\setcounter{tocdepth}{1}

\title[DR6 Lensing foregrounds]{The Atacama Cosmology Telescope: Mitigating the impact of extragalactic foregrounds for the DR6 CMB lensing analysis}
 \shorttitle{ACT DR6 Lensing Foregrond Mitigation}
  \shortauthors{MacCrann, Sherwin, Qu, Namikawa, Madhavacheril et al.}

\author{Niall~MacCrann}\affiliation{DAMTP, Centre for Mathematical Sciences, University of Cambridge, Wilberforce Road, Cambridge CB3 OWA, UK}
\author{Blake~D.~Sherwin}\affiliation{DAMTP, Centre for Mathematical Sciences, University of Cambridge, Wilberforce Road, Cambridge CB3 OWA, UK}\affiliation{Kavli Institute for Cosmology Cambridge, Madingley Road, Cambridge CB3 0HA, UK}
\author{Frank~J.~Qu}\affiliation{DAMTP, Centre for Mathematical Sciences, University of Cambridge, Wilberforce Road, Cambridge CB3 OWA, UK}
\author{Toshiya~Namikawa}\affiliation{Kavli IPMU (WPI), UTIAS, The University of Tokyo, Kashiwa, 277-8583, Japan}
\author{Mathew~S.~Madhavacheril}\affiliation{Department of Physics and Astronomy, University of
Pennsylvania, 209 South 33rd Street, Philadelphia, PA, USA 19104}
\author{Irene~Abril-Cabezas}\affiliation{DAMTP, Centre for Mathematical Sciences, University of Cambridge, Wilberforce Road, Cambridge CB3 OWA, UK}
\author{Rui~An}\affiliation{Department of Physics and Astronomy, University of Southern California, Los Angeles, CA 90089, USA}
\author{Jason~E.~Austermann}\affiliation{NIST, Quantum Sensors Group, 325 Broadway, Boulder, CO, 80305}
\author{Nicholas~Battaglia}\affiliation{Department of Astronomy, Cornell University, Ithaca, NY 14853, USA}
\author{Elia~S.~Battistelli}\affiliation{Sapienza—University of Rome—Physics department, Piazzale Aldo Moro 5—I-00185, Rome, Italy
}
\author{James~A.~Beall}\affiliation{NIST, Quantum Sensors Group, 325 Broadway, Boulder, CO, 80305}
\author{Boris~Bolliet}\affiliation{DAMTP, Centre for Mathematical Sciences, University of Cambridge, Wilberforce Road, Cambridge CB3 OWA, UK}
\author{J.~Richard~Bond}\affiliation{Canadian Institute for Theoretical Astrophysics, University of Toronto, Toronto, ON, Canada M5S 3H8}
\author{Hongbo~Cai
}\affiliation{Department of Physics and Astronomy, University of Pittsburgh, Pittsburgh, PA, USA 15260}
\author{Erminia~Calabrese}\affiliation{School of Physics and Astronomy, Cardiff University, The Parade, Cardiff, Wales CF24 3AA, UK}
\author{William~R.~Coulton}\affiliation{Center for Computational Astrophysics, Flatiron Institute, 162 5th Avenue, New York, NY 10010, USA}
\author{Omar Darwish}\affiliation{Université de Genève, Département de Physique Théorique and Centre for Astroparticle Physics, 24 quai Ernest-Ansermet, CH-1211 Genève 4, Switzerland}
\author{Shannon~M.~Duff}\affiliation{NIST, Quantum Sensors Group, 325 Broadway, Boulder, CO, 80305}
\author{Adriaan~J.~Duivenvoorden}\affiliation{Center for Computational Astrophysics, Flatiron Institute, 162 5th Avenue, New York, NY 10010, USA}\affiliation{Joseph Henry Laboratories of Physics, Jadwin Hall, Princeton University, Princeton, NJ, USA 08544}
\author{Jo~Dunkley}\affiliation{Joseph Henry Laboratories of Physics, Jadwin Hall, Princeton University, Princeton, NJ, USA 08544}
\affiliation{Department of Astrophysical Sciences, Peyton Hall,  Princeton University, Princeton, NJ 08544, USA}
\author{Gerrit~S.~Farren}\affiliation{DAMTP, Centre for Mathematical Sciences, University of Cambridge, Wilberforce Road, Cambridge CB3 OWA, UK}
\author{Simone~Ferraro}\affiliation{Lawrence Berkeley National Laboratory, One Cyclotron Road, Berkeley, CA 94720, USA}
\author{Joseph~E.~Golec}
\affiliation{Department of Physics, University of Chicago, 5720 South Ellis Avenue, Chicago, IL 60637, USA}
\affiliation{Kavli Institute for Cosmological Physics, University of Chicago, 5640 South Ellis Avenue, Chicago, IL 60637, USA}
\author{Yilun~Guan}
\affiliation{Dunlap Institute for Astronomy and Astrophysics, University of Toronto, 50 St. George St., Toronto, ON M5S 3H4, Canada}
\author{Dongwon Han}\affiliation{DAMTP, Centre for Mathematical Sciences, University of Cambridge, Wilberforce Road, Cambridge CB3 OWA, UK}
 \author{Carlos Herv\'ias-Caimapo} \affiliation{Instituto de Astrof\'isica and Centro de Astro-Ingenier\'ia, Facultad de F\'isica, Pontificia Universidad Cat\'olica de Chile, Av. Vicu\~na Mackenna 4860, 7820436 Macul, Santiago, Chile}
\author{J.~Colin Hill}\affiliation{Department of Physics, Columbia University, 538 West 120th Street, New York, NY, USA 10027}
\author{Matt~Hilton}\affiliation{Wits Centre for Astrophysics, School of Physics, University of the Witwatersrand, Private Bag 3, Johannesburg 2050, South Africa}
\author{Ren\'ee~Hlo\v{z}ek}\affiliation{David A. Dunlap Department of Astronomy \& Astrophysics, 50 St George Street, Toronto ON M5S 3H4}
\affiliation{Dunlap Institute for Astronomy \& Astrophysics,  50 St George Street, Toronto ON M5S 3H4}
\affiliation{Astrophysics Research Centre, University of KwaZulu-Natal, Westville Campus, Durban 4041, South Africa}
\author{Johannes~Hubmayr}\affiliation{NIST, Quantum Sensors Group, 325 Broadway, Boulder, CO, 80305}
\author{Joshua~Kim}\affiliation{Department of Physics and Astronomy, University of Pennsylvania, 209 South 33rd Street, Philadelphia, PA, USA 19104}
\author{Zack Li 
}
\affiliation{Canadian Institute for Theoretical Astrophysics, University of Toronto, Toronto, ON, Canada M5S 3H8}
\author{Arthur~Kosowsky}\affiliation{Department of Physics and Astronomy, University of Pittsburgh, Pittsburgh, PA 15260 USA}
\author{Thibaut~Louis}\affiliation{Université Paris-Saclay, CNRS/IN2P3, IJCLab, 91405 Orsay, France}
\author{Jeff~McMahon}\affiliation{Kavli Institute for Cosmological Physics, University of Chicago, 5640 South Ellis Avenue, Chicago, IL 60637, USA}
\affiliation{Department of Astronomy and Astrophysics, University of Chicago, 5640 S. Ellis Ave., Chicago, IL 60637, USA}
\affiliation{Department of Physics, University of Chicago, 5720 South Ellis Avenue, Chicago, IL 60637, USA}
\affiliation{Enrico Fermi Institute, University of Chicago, Chicago, IL 60637, USA}
\author{Gabriela~A.~Marques}\affiliation{Fermi National Accelerator Laboratory, P. O. Box 500, Batavia, IL 60510, USA}
\author{Kavilan~Moodley}\affiliation{Astrophysics Research Centre, School of Mathematics, Statistics and Computer Science, University of KwaZulu-Natal, Durban 4001, South Africa}
\author{Sigurd~Naess}\affiliation{Institute of theoretical
astrophysics, University of Oslo, Norway
}
\author{Michael~D.~Niemack}\affiliation{Department of Physics, Cornell University, Ithaca, NY 14853, USA}
\affiliation{Department of Astronomy, Cornell University, Ithaca, NY 14853, USA}
\author{Lyman Page}\affiliation{Joseph Henry Laboratories of Physics, Jadwin Hall, Princeton University, Princeton, NJ, USA 08544}
\author{Bruce~Partridge}\affiliation{Department of Physics and Astronomy, Haverford College, Haverford PA, USA 19041}
\author{Emmanuel Schaan}
\affiliation{SLAC National Accelerator Laboratory, Menlo Park, CA 94025, USA}
\affiliation{Kavli Institute for Particle Astrophysics and Cosmology and Department of Physics, Stanford University, Stanford, CA 94305, USA}
\author{Neelima~Sehgal}\affiliation{Physics and Astronomy Department, Stony Brook University, Stony Brook, New York 11794, USA}
\author{Crist\'obal Sif\'on}\affiliation{Instituto de F\'isica, Pontificia Universidad Cat\'olica de Valpara\'iso, Casilla 4059, Valpara\'iso, Chile}
\author{Edward~J.~Wollack}\affiliation{NASA Goddard Space Flight Center, 8800 Greenbelt Rd, Greenbelt, MD 20771 USA}
\author{Maria~Salatino}\affiliation{Physics Department, Stanford University, 382 via Pueblo, Stanford 94305 CA USA}
\author{Joel~N.~Ullom}\affiliation{NIST, Quantum Sensors Group, 325 Broadway, Boulder, CO, 80305}
\author{Jeff~Van~Lanen}\affiliation{NIST, Quantum Sensors Group, 325 Broadway, Boulder, CO, 80305}
\author{Alexander~Van~Engelen}\affiliation{School of Earth and Space Exploration, Arizona State University, Tempe, AZ, USA 85287}
\author{Lukas~Wenzl}\affiliation{Department of Astronomy, Cornell University, Ithaca, NY 14853, USA}







\begin{abstract}
We investigate the impact and mitigation of extragalactic foregrounds for the CMB lensing power spectrum analysis of Atacama Cosmology Telescope (ACT) data release 6 (DR6) data. Two independent microwave sky simulations are used to test a range of mitigation strategies. We demonstrate that  finding and then subtracting point sources, finding and then subtracting models of clusters, and using a profile bias-hardened lensing estimator, together reduce the fractional biases to well below statistical uncertainties, with  the inferred lensing amplitude, $\Alens$,  biased by less than $0.2\sigma$. We also show that another method where a model for the cosmic infrared background (CIB) contribution is deprojected and high frequency data from \planck\ is included has similar performance. Other frequency-cleaned options do not perform as well, incurring either a large noise cost, or resulting in biased recovery of the lensing spectrum. In addition to these simulation-based tests, we also present null tests performed on the ACT DR6 data which test for sensitivity of our lensing spectrum estimation to differences in foreground levels between the two ACT frequencies used, while nulling the CMB lensing signal. These tests pass whether the nulling is performed at the map or bandpower level. The CIB-deprojected measurement performed on the DR6 data is consistent with our baseline measurement, implying contamination from the CIB is unlikely to significantly bias the DR6 lensing spectrum. This collection of tests gives  confidence that the ACT DR6 lensing measurements and cosmological constraints presented in companion papers to this work are robust to extragalactic foregrounds. 

\end{abstract}



\section{Introduction}

Gravitational lensing provides a relatively direct method of probing the matter distribution in the universe, an otherwise challenging task given the domination of dark matter over visible matter. By measuring lensing statistics we can therefore constrain the properties of dark energy and neutrinos by modeling their impact on the statistics of the matter distribution across a wide range in redshift. The cosmic microwave background (CMB) is a useful source of photons for lensing since its statistics in the absence of lensing are close to Gaussian and isotropic. Furthermore, we know its redshift precisely, which is required for inference since the strength of lensing depends on source redshift. The deflection of CMB photons by lensing, induced by a given realization of the matter field, breaks the statistical isotropy of the CMB, generating an off-diagonal covariance between Fourier modes.
This  can be used to construct quadratic estimators for the lensing convergence which is closely related to the projected matter density field (see \citealt{lewis2006} for a review of CMB lensing). 

One of the main challenges in robust CMB lensing estimation comes from the presence of other \emph{secondary} anisotropies, i.e. physical processes between us and the surface of last scattering that generate millimetre radiation (such as emission from dusty galaxies and radio sources) or scatter the CMB photons, such as the thermal and kinematic Sunyaev-Zeldovich effects (henceforth tSZ and kSZ respectively). These sources of statistical anisotropy can bias the lensing reconstruction.
Fortunately, various methods have been developed that mitigate these effects. 

Firstly, the frequency dependence of these foregrounds can be exploited to project out a given spectral energy distribution (SED) or equivalently, form linear combinations of individual frequency maps that null a given SED (e.g.  \citealt{remazeilles11,madhavacheril18}). High signal-to-noise point sources and clusters can be detected using a matched-filter approach (e.g. \citealt{haehnelt2006,Staniszewski2009}), for which models can be fitted and subtracted, or regions around these detections can be masked or ``inpainted'' (e.g. \citealt{bucher12}). All of these methods aim to remove the contaminating contributions of foregrounds to the maps entering the lensing reconstruction. 

Secondly, there are methods developed for lensing estimation specifically that amend the usual quadratic estimator for lensing reconstruction to reduce sensitivity to foregrounds. The quadratic estimator reconstructs lensing potential modes, $\phi(\vecL)$, by averaging over pairs of temperature modes, $T(\vecl)T(\vecL-\vecl)$. \emph{Bias-hardening} \citep{osborne14, namikawa14,sailer20} involves amending the estimator such that it has zero response to the mode coupling generated by, for example,  Poisson distributed sources. Finally, the \emph{shear} estimator has recently been developed by \cite{schaan19} (and generalized to the curved-sky by \citealt{qu22}). This estimator uses only the quadrupolar contribution to the coupling between observed CMB Fourier modes induced by lensing, which is largely unaffected by extragalactic foregrounds \citep{schaan19}.

We note here that extragalactic rather than Galactic foregrounds are the focus of this paper. Galactic dust contamination falls sharply at small scales (high $l$), and the accompanying lensing power spectrum estimation in \cite{dr6-lensing-auto} uses CMB modes at $l>600$ only, which should ensure minimal contamination from Galactic foregrounds \citep{challinor18,beck20}. \cite{dr6-lensing-auto} also demonstrate that the measured lensing power spectrum is stable to using an even more conservative $l_{\mathrm{min}}$. Furthermore, unlike extragalactic foregrounds, we can use the large-scale anisotropy of Galactic foregrounds (they are much higher in amplitude close to the Galactic plane) to test sensitivity. If the lensing power spectrum measurement was significantly biased by Galactic foregrounds, that bias would be highly sensitive to the strictness of the Galactic mask used, however, \cite{dr6-lensing-auto} demonstrate that the measured lensing power spectrum is very stable to using a more conservative Galactic mask than the baseline choice. 

In recent years the \planck\ satellite has provided the data for the state-of-the-art CMB lensing reconstruction and (auto-) power spectrum measurements \citep{plancklensing2018, carron22}, building on initial detections of CMB lensing cross-correlations with \textit{WMAP} satellite data \citep{smith07,hirata08}, and the lensing (auto) power spectrum by the ground-based Atacama Cosmology Telescope\footnote{\url{https://act.princeton.edu/}} (ACT) and the South Pole Telescope\footnote{\url{https://pole.uchicago.edu/public/Home.html}} (SPT) \citep{das11,  vanengelen12}.
Much larger telescopes and detector arrays are feasible for ground-based experiments, allowing for higher resolution, lower noise observations. With enhanced instrumentation, ACT and SPT data can now achieve comparable statistical power to \planck. Meanwhile the upcoming Simons Observatory\footnote{\url{https://simonsobservatory.org/}} (SO), and the planned CMB-S4\footnote{https://cmb-s4.org/} \citep{Abazajian16} will further increase  sensitivity, especially in polarisation. 

However, these data come with challenges in controlling foregrounds. Firstly, the higher CMB $l$-modes (i.e. smaller angular scales) accessible with these experiments are more contaminated by certain extragalactic foregrounds, especially dusty galaxies and the Sunyaev-Zeldovich effects (see e.g. \citealt{vanengelen14, osborne14}). Ground-based experiments also have reduced frequency coverage at these high CMB multipoles compared to that leveraged by \planck\ to perform multi-frequency cleaning \citep{planck2018maps}; for this analysis we use only f090 (77-112 GHz) and f150
(124-172 GHz) frequency bands from ACT.

Despite these challenges, we argue in the following that we can effectively mitigate such extragalactic foregrounds with ACT data release 6 (DR6) data, enabling robust lensing power spectrum measurements and cosmological inference, as described in companion papers \citet{dr6-lensing-auto} and \citet{dr6-lensing-cosmo}. In \sect{sec:formalism}, we briefly cover relevant formalism for lensing quadratic estimators (including bias-hardening and frequency-cleaned estimators), and for quantifying biases from extragalactic foregrounds. In \sect{sec:actstuff}, we describe the ACT DR6 data and give a brief overview of the DR6 lensing analysis of \cite{dr6-lensing-auto} to provide context for the subsequent foreground bias estimates. In \sect{sec:simulatiionprc}, we describe the microwave-sky simulations used to test our mitigation strategies, and in \sect{sec:simpredictions} we present predicted biases from these simulations. In \sect{sec:nulls}, we present null tests performed on the DR6 data, which test for sensitivity to the difference in foreground biases for the two ACT frequency channels used, and present consistency of a CIB-deprojected analysis with our baseline analysis. We present our conclusions in \sect{sec:conclusion}.

\section{Formalism and Methods}\label{sec:formalism}

We briefly describe the formalism associated with the (bias-hardened) quadratic estimators and multi-frequency-cleaning approaches used in this work. Note that we focus on the CMB temperature field, since we assume polarisation is not affected by extragalactic foregrounds - certainly we do not expect fields such as the thermal Sunyaev-Zeldovich effect (tSZ) and cosmic infrared background (CIB) to be 
polarised at a significant level for our observations, and any bright polarised sources in the DR6 data are subtracted or masked and inpainted \citep{dr6-lensing-auto, dr6-maps}.
\subsection{Quadratic Estimator}\label{sec:qe}

We use throughout the quadratic estimator \citep{huokamoto2002}, using the curved-sky calculations implemented in \textsc{falafel}\footnote{\url{https://github.com/simonsobs/falafel}}, and the normalization and reconstruction noise calculations implemented in \textsc{tempura}\footnote{\url{https://github.com/simonsobs/tempura}}.  For simplicity, we will present in the text the flat-sky equations, where the quadratic estimator for Fourier mode $L$ of the lensing potential,  $\phi_{\vecL}$, is given by (e.g. \citealt{hanson11})
\begin{equation}
    \hat{\phi}_{\vecL} = \frac{1}{2} \Aphi_{\vecL} \int \frac{\dx{^2\vecl}}{(2\pi)^2}  \frac{f_{\vecL,\vecl} T_{\vecl}T_{\vecL-\vecl}}{\Cltot{l}\Cltot{|\vecL-\vecl|}},
    \label{eq:qe}
\end{equation}
with the normalisation factor $\Aphi_{\vecL}$ given by
\begin{equation}
    \Aphi_{\vecL} = \left[ \int \frac{\dx{^2\vecl}}{(2\pi)^2} \frac{f_{\vecL,\vecl}^2}{2\Cltot{l}\Cltot{|\vecL-\vecl|}} \right]^{-1}
    \label{eq:norm}
\end{equation}
and 
\begin{equation}
    f_{\vecL,\vecl} = \tilde{C}_l\vecl \cdot \vecL + \tilde{C}_{|\vecL-\vecl|}\vecL \cdot |\vecL-\vecl|.
    \label{eq:fphi}
\end{equation}
Here $\tilde{C}_l$ is the lensed CMB temperature power spectrum, while $\Cltot{l}$ is the total observed CMB power spectrum (i.e. including noise). It is also common to work with the convergence, $\kappa$ (and its power spectrum $\Clkappa{L}$), instead of the lensing potential, which is simply related to $\phi$ via
\begin{equation}
    \kappa_{\vecL} = L(L+1) \phi_{\vecL} / 2.
\end{equation}

The power spectrum of the reconstruction has what is known as an $N^0$ bias, which is the expectation for the quadratic estimator operating on a Gaussian, statistically isotropic field with the same power spectrum as the lensed CMB (i.e. in the absence of the statistcal anisotropy induced by lensing). For the basic, temperature-only, quadratic estimator described here, $N^0$ is equal to $A^{\phi}_{\vecL}$. $N^0$ can be calculated analytically or from simulations, and subtracted off. The realization-dependent $N^0$ (RDN0), method of \citet{hanson11, namikawa13} minimizes the impact of small differences between the CMB power spectrum assumed in the $N^0$ calculation, and the true CMB power spectrum in the data.

\subsection{Foreground biases}\label{sec:fg_biases}

Foregrounds perturb the CMB temperature, $\Tcmb$, as 
\begin{equation}
    T(\vecl) = \Tcmb(\vecl) + \Tfg(\vecl).
\end{equation}
Denoting a quadratic estimator (see \eqn{eq:qe}) on two temperature maps $A$ and $B$ as $Q(T^A, T^B)$, the contamination of $C_L^{\hat{\phi}\hat{\phi}}$ is (e.g. \citealt{vanengelen14, osborne14})
\begin{equation}
\begin{split}
    \Delta \Clhatphi{L}
    &= 2\left<Q(\Tfg,\Tfg) \phi \right>_L\\
    &+ 4\left<Q(\Tfg, \Tcmb)Q(\Tfg, \Tcmb)\right>_L\\
    &+ \left<Q(\Tfg,\Tfg)Q(\Tfg,\Tfg)\right>_L.
\end{split}
\label{eq:fg_expansion}
\end{equation}
where we have made the subtitution $Q(\Tcmb,\Tcmb)=\phi$ in the first line.
The first and second terms are known as the \emph{primary} and \emph{secondary} \emph{bispectrum} terms respectively, since they depend on bispectra involving $\Tfg$ and $\phi$ (in the secondary bispectrum case the $\phi$-dependent term arises from the presence of one $\Tcmb$ in each of the quadratic estimators).  The third term is known as the trispectrum contribution, since it depends only on the trispectrum of $\Tfg$. Given a simulation of $\Tfg$ and $\phi$, we can estimate these terms individually, and without noise from the primary CMB, following \citet{schaan19}, and briefly summarized here: The primary bispectrum term is computed directly as in the first line of \eqn{eq:fg_expansion}, using the true $\phi$-field for the simulation. Noise on the secondary bispectrum estimate is reduced greatly by using the difference $\left<Q(\Tfg, \Tcmb)Q(\Tfg, \Tcmb)\right>_L - \left<Q(\Tfg, \Tcmb')Q(\Tfg, \Tcmb')\right>_L$, where $\Tcmb'$ is formed from the same unlensed CMB as $\Tcmb$, but lensed by an independent $\kappa$ field.
The estimator for the trispectrum term in \eqn{eq:fg_expansion} has an ``$N^0$ bias" (the disconnected trispectrum) that we subtract\footnote{Note in a real data analysis this would be accounted for by RDN0 correction described above, in  the usual case where, after some foreground mitigation, foreground power is sub-dominant to CMB power.}; this is given by 
\begin{equation}
    N^{0,\mathrm{fg}}_{L} =  (\Aphi_{\vecL})^2 \int \frac{\dx{^2\vecl}}{(2\pi)^2}  \frac{f_{\vecL,\vecl}^2 C^{\mathrm{fg}}_l C^{\mathrm{fg}}_{|\vecL-\vecl|}}{2(C_l^{\mathrm{tot}}C_{|\vecL-\vecl|}^{\mathrm{tot}})^2}.
    \label{eq:N0fg}
\end{equation}
Note that the integral here is the same as for $\Aphi_L$, but replacing $\Cltot{l}$ with $(\Cltot{l})^2 / \Clfg{l}$.

\subsection{Bias-hardened estimators}\label{sec:bh}

In general a bias-hardened estimator for a field $x_\vecL$ in the presence of a contaminant $y_\vecL$ is \citep{namikawa13,osborne14,sailer20}
\begin{equation}
    x_\vecL = \frac{\hat{x}_{\vecL}-A^x_{\vecL} R^{xy}_{\vecL}\hat{y}_\vecL}{1-A^x_{\vecL}N^{y}_{\vecL} (R^{xy}_{\vecL})^2}
\end{equation}
where $\hat{x}_\vecL$ is the non-hardened estimator, given by
\begin{equation}
\hat{x}_\vecL =     \frac{1}{2} A^{x}_{\vecL} \int \frac{\dx{^2\vecl}}{(2\pi)^2}  \frac{f^{x}_{\vecL,\vecl} T_{\vecl}T_{\vecL-\vecl}}{\Cltot{l}\Cltot{|\vecL-\vecl|}}
\end{equation} 
and $\hat{y}_\vecL$ is defined analogously. 
$R^{xy}$ is the response of the estimator for field $x$ to the presence of field $y$, and is given by
\begin{equation}
    R_{\vecL}^{xy} = \int \frac{\dx{^2\vecl}}{(2\pi)^2} \frac{f^{y}_{\vecL,\vecl} f^{x}_{\vecL,\vecl}}{2\Cltot{l}\Cltot{|\vecL-\vecl|}}.
\end{equation}
$A^{x}_{\vecL}$ is the estimator normalization, given by $1/R_{\vecL}^{xx}$. Here $f^{x}$ and $f^{y}$ are functions which describe the mode-coupling induced in the observed CMB by the fields $x$ and $y$. 
For lensing, this is given by \eqn{eq:fphi} while for point-sources it is a constant. \citet{sailer20} showed that this method can also harden against Poisson distributed sources with some (Fourier space) radial profile $u(\vecl)$ (with unknown amplitude), in this case, 
\begin{equation}
    f_{\vecl,\vecL-\vecl} = \frac{u(\vecl)u(\vecL-\vecl)}{u(\vecL)}.
\end{equation}
\citet{sailer23b} further demonstrate that good performance can be achieved even with an imperfect choice profile (e.g. CIB is mitigated even when using a profile chosen for tSZ). 

Since we are effectively deprojecting modes that contain some information on lensing, there is a noise cost to doing bias-hardening (for the DR6 lensing analysis presented in \citet{dr6-lensing-auto} it is $\sim10\%$), with the $N^0$ of the bias-hardened estimator given by
\begin{equation}
    N^{0,x,BH} = N^{0,x} / (1 - N^{0,x} N^{0,y} (R^{xy})^2).
\end{equation}

When computing the foreground trispectrum contamination, we again need to subtract an $N^{0}$ contribution (via the last term in \eqn{eq:fg_expansion}), given by
\begin{equation}
    N^{0,\mathrm{fg,BH}} = 
\frac{
    N^{0,\mathrm{fg}} + (A^x)^2 (R^{xy}A^{y} - 2  R^{xy} A^{y}  R^{xy}_{\mathrm{fg}})}{\left[1 - A^x A^y (R^{xy})^2\right]^2}
\end{equation}
where $N^{0,\mathrm{fg}}$ is given by \eqn{eq:N0fg} (and is also equal to $(A^x_{\vecL})^2 / R^{xx,\mathrm{fg}}_{\vecL}$),  
\begin{equation}
    R^{xy,\mathrm{fg}} = \int \frac{\dx{^2\vecl}}{(2\pi)^2} \frac{f^{y}_{\vecL,\vecl} f^{x}_{\vecL,\vecl} \Clfg{l}\Clfg{|\vecL-\vecl|}}{2\Cltot{l}\Cltot{|\vecL-\vecl|}}
\end{equation}
 and $\Clfg{l}$ is the power spectrum of the foreground-only temperature field. 

\subsection{Frequency-cleaned, asymmetric and  symmetrised estimators}\label{sec:theoryfreqclean}

Even for the temperature-only case, we need not use the same two maps in our quadratic estimator for the lensing potential; we can also use two temperature maps which have had different levels of foreground cleaning applied, which can result in a better bias-variance trade-off than using the same (noisy) frequency-cleaned temperature map in both legs of the quadratic estimator. Indeed, if just one of the maps has zero foreground contamination, we might expect the quadratic estimator to be unbiased by foregrounds in cross-correlation with the true convergence field or some tracer of it (see \citealt{hu07, madhavacheril18, darwish21}). However, the secondary bispectrum contribution to the auto-power spectrum of the reconstruction is not removed in general.

The standard quadratic estimator in \eqn{eq:qe} is not symmetric when two different temperature maps are used i.e. $\phi_{XY} = Q\left(X,Y\right)$ is not in general equal to $\phi_{YX} = Q\left(Y,X\right)$, for $X\neq Y$. \citet{darwish21a} therefore propose using a minimum-variance linear combination of the two asymmetric estimators:
\begin{equation}
    \phi_{\mathrm{sym}}(\vecL) = \bm{W}(\vecL) \begin{pmatrix}
    \phi_{XY}(\vecL) \\
    \phi_{YX}(\vecL)
    \end{pmatrix}
\end{equation}
where $\bm{W}(\vecL)$ is some weight matrix. We define the matrix 
\begin{equation}
    \textbf{N}^0(\vecL) = \begin{pmatrix}
N^{0,XYXY}(\vecL) & N^{0,XYYX}(\vecL) \\
N^{0,YXXY}(\vecL) & N^{0,YXYX}(\vecL)
\end{pmatrix}
\end{equation}
where $N^{0,ABCD}(\vecL)$ is the variance of the lensing reconstruction power spectrum for the general case of the cross-correlation of lensing reconstructions $\phi_{AB}$ and $\phi_{CD}$, derived from CMB temperature maps $A,B,C,D$ (see end of \sect{sec:qe} for discussion of $N^0$).

Then the normalized minimum variance $\bm{W}(\vecL)$ is given by
\begin{equation}
W(\vecL) = \frac{\left(\textbf{N}^0\right)^{-1}}{|\textbf{N}^0|(N^{0,XYXY}-N^{0,YXYX})}
\end{equation}
where we have used that $N^{0,XYYX}=N^{0,YXXY}$. We'll refer to this as the \emph{D21} estimator.

\section{ACT DR6 data and lensing power spectrum analysis}
\label{sec:actstuff}

We  briefly describe the ACT DR6 data and how it is used for the lensing power spectrum analysis described in \citet{dr6-lensing-auto,dr6-lensing-cosmo}, including summarizing the baseline foreground mitigation, which will provide useful context for our description of the simulation processing in \sect{sec:actsim}.

ACT DR6 includes maximum-likelihood maps at $0.5^\prime$ resolution. The lensing power spectrum measurement \citep{dr6-lensing-auto} which this paper aims to support  uses the first science-grade version of the ACT DR6 maps, labeled dr6.01 and generated from observations performed between May 2017 and June 2021\footnote{Since these maps were generated, the ACT team have made some refinements to the mapmaking that improve the large-scale transfer function and polarisation noise levels, and include data taken in 2022. The team expect to use a second version of the maps for the DR6 public data release, and for further lensing analyses.}. For the lensing analysis, we use the f090 band and f150 band data (with central frequencies $\sim97$ and $\sim150$ GHz respectively).
For each frequency, for temperature (T) and polarisation data (Q and U), an optimal coadd (including data taken during the daytime not used in the lensing analysis) is generated, and a point-source catalog is generated using a matched-filter algorithm. These point-source catalogs are used to subtract point-source models from each single-array map\footnote{i.e. each map corresponding to a portion of the data observed by a given detector array at a certain frequency} before they enter a weighted coaddition step in harmonic space, with weights given by the inverse noise power spectrum of a given map (see Section 4.6 of  \citet{dr6-lensing-auto} for further details). In addition, for the temperature data in our baseline approach, we subtract a map that models the tSZ contributions from detected clusters, which is a superposition of the best-fit templates inferred by the matched-filter cluster finding code \textsc{nemo}\footnote{\url{https://nemo-sz.readthedocs.io/en/latest/}} \citep{Hilton_2021}, for $S/N>5$ detections (see \citealt{Hilton_2021} and Section 2.1.3 of \citealt{dr6-lensing-auto}).

The lensing reconstruction and power spectrum estimation is performed using the  ``cross-correlation-only'' estimator of \citet{madhavacheril20b}, with our implementation using maps constructed from four independent splits of the DR6 data. The final lensing power spectrum estimate is constructed by combining the various different estimators one can form from using a different one of the four independent data splits in each of the four legs of the lensing power spectrum estimator (see Section 4.8.1 of \citealt{dr6-lensing-auto}). We use a \emph{`minimum variance'}\footnote{this is in quotes because our implementation does not take into account all cross-correlations between pairs AB in \eqn{eq:phimv}, see \citet{dr6-lensing-auto} for more details.} (henceforth `MV') lensing reconstruction that combines temperature and polarisation information via a weighted sum of the reconstructions from all pairs $XY \in \left[TT,TE,TB,EE,EB\right]$:
\begin{equation}
    \hat{\phi}^{\mathrm{MV}}(\vecL) = \sum_{XY} w_{XY} \hat{\phi}^{AB}(\vecL)
    \label{eq:phimv}
\end{equation}
with weights given by the inverse of the normalization (\eqn{eq:norm} for the $TT$ case).
. 

Through the tests described in this paper, we arrive at a set of foreground mitigation analysis choices for the baseline ACT DR6 lensing analysis, which we summarize briefly here:
\begin{enumerate}
    \item Models for $S/N>4$ point sources are subtracted from each map entering the coadd. This helps mitigate the impact of radio point sources and the brighter among the dusty galaxies that constitute the CIB. 
    \item A cluster tSZ model map, generated from cluster candidates identified by the \nemo\ code, is subtracted from each frequency map (see section \sect{sec:actsim} for more details).
    \item A bias-hardened estimator is used for the TT contribution to the $\hat{\phi}^{\mathrm{MV}}(\vecL)$ estimate, with profile proportional to $(C_l^{tSZ})^{0.5}$, where $C_l^{tSZ}$ is the tSZ power spectrum, which we estimate from the \websky\ simulations (see \sect{sec:bh} for discussion of bias-hardening). We will refer to this estimator as the \emph{profile-hardened estimator}.
\end{enumerate}

\section{Simulations and processing}
\label{sec:simulatiionprc}

In this section, we use two extragalactic microwave sky simulations (described in \sect{sec:simulations}), to make predictions for biases caused by contamination due to extragalactic foregrounds to the estimated lensing power spectrum. 

\subsection{Microwave-sky simulations}\label{sec:simulations}

Given the multiple complex astrophysical processes which distort or contaminate the observed CMB, we start by predicting biases in the estimated lensing power spectrum from cosmological simulations which attempt to model the nonlinear, correlated fields that constitute extragalactic foregrounds. We use two such simulations: the \websky\ simulations \citep{websky}, and the simulations from \citet{Sehgal_2010}, which we will refer to as the \emph{S10} simulations. 

The \websky\ simulation is built upon a full-sky dark matter halo catalog based on a $600(\text{Gpc}/h)^3$ mock matter field generated using the fast ``mass-Peak Patch'' algorithm \citep{stein19}. CIB is generated using a halo model with parameters fit to Herschel power spectrum measurements (see \citealt{shang12,viero13} for details). Gas is distributed around halos by assuming spherically symmetric, parametric radial distributions from \citet{battaglia12}, which are then used to model the kinetic Sunyaev-Zeldovich effect, and the Compton-$y$ field. Maps of radio emission from galaxies are also generated using a halo model, as described in \citet{Li22}.

The \sehgal\ simulation is built upon a 1$h^{-1}$Gpc N-body simulation, with lensing quantities calculated via ray-tracing. Gas is added to the dark-matter halos without assuming spherical symmetry (following \citealt{bode2007}), allowing for additional complexity in the SZ observables relative to \websky. Radio sources and infrared galaxies are added using halo occupation distributions tuned to a variety of multi-wavelength observations (see \citealt{Sehgal_2010} for details).

\subsection{Simulated ACT DR6 maps}\label{sec:actsim}

In order to generate realistic estimates of the contamination of the ACT DR6 lensing power spectrum due to extragalactic foregrounds, we need to perform some pre-processing of the microwave sky simulations described in \sect{sec:simulations}, such that they have some of the same observational properties as the real ACT data. 

\begin{enumerate}
\item{We generate (in differential CMB temperature units), a total foreground map (for each frequency) by summing the contributions from the tSZ, kSZ, CIB and radio point-sources. We neglect the dependence of the foreground components across the ACT passbands (since we only have simulation outputs at a small number of discrete single frequencies), and simply choose the  simulated frequency closest in frequency to  90 and 150 GHz (93 and 145 GHz for \websky, 90 and 148 GHz for \sehgal). This is a reasonable approach given that there is theoretical uncertainty in the foreground components studied here that is larger than the  $\mathcal{O}(10\%)$ errors induced by neglecting the passbands. Fortunately, as we will show, the residual biases (as a fraction of the lensing signal) after foreground mitigation are at the percent level, hence the dependence across passbands is only relevant at the $\mathcal{O}(0.1\%)$, well below our current statistical precision.}

\item{We then add a realization of the CMB, lensed by the appropriate $\kappa$ field.}
\item{We then convolve with a Gaussian beam appropriate for the frequency (with full-width-at-half-maximum (FWHM) = 2.2 and 1.4 arcminutes for 90 and 150 GHz respectively), convert the maps to the \textit{Plate-Carr\'{e}e} cylindrical (CAR) pixelization used for ACT data and pipelines, and add a random  realization of the instrumental noise, approximated as being  Gaussian with the spatially varying inverse variance estimated for the DR6 day+night coadded data\footnote{Note that this is the noise level of the maps used for source and cluster finding, while the DR6 lensing analysis uses only night-time observations, which have higher noise. However, the method used here for calculating foreground biases does not require simulating that latter noise level.}  (see \citealt{dr6-maps}).}
\item{We input these simulated maps to the matched-filter source and cluster finding algorithms implemented by \textsc{nemo}. For each frequency, we initially run \nemo\ in point-source finding mode with a $\snr>4$ threshold, which outputs a point-source catalog which can be used to generate a point-source model map.}
\item{We then run \nemo\ in cluster-finding mode, jointly on the 90 and 150 GHz maps. The matched-filter templates are a range of cluster model templates, in this case based on the Universal Pressure Profile of \citet{arnaud10}\footnote{We note here that this is not the same theoretical model for the tSZ profile used in constructing the \websky\ simulations (which used \citet{battaglia12}); the fact that we find tSZ biases in \websky\ are well under control encourages us that our foreground biases are not very sensitive to the theoretical cluster profile assumed when detecting and subtracting models for clusters.}. 15 filters with different angular sizes are constructed by varying the mass and redshift of the cluster model (see \citet{Hilton_2021} for details).  
The point-source catalogs from the previous steps are used to mask point-sources during the cluster finding. This outputs a catalog of candidate clusters with estimated $S/N$ and best-fit cluster model template (with the best-fit redshift and halo mass simply that of the best-fit template).}
\end{enumerate}

For the baseline DR6 lensing analysis, models for both the point-source and tSZ cluster contribution are subtracted at the map level. 
We therefore perform this model subtraction on the foreground-only maps that are required for the foreground bias estimator used here (described in \sect{sec:fg_biases}).


We present in the appendices two variations on this procedure to address concerns about its accuracy. First, for both simulations, we have limited volume, so the bias estimates we obtain may contain significant statistical uncertainty due to the limited area of the foreground-only maps used; especially after applying the $f_{\mathrm{sky}}\sim0.3$ DR6 analysis mask. In \app{app:webskyrot} we present foreground bias estimates from a  version of the \textsc{websky} simulations rotated such that a non-overlapping region of the simulation populates the DR6 mask. The foreground biases are consistent, implying that cosmic variance is not an important contributor to uncertainty in the foreground bias estimates presented here. 

Second, we note that the noise model assumed above is rather simplified; while it probably includes  the small-scale noise important for cluster and source-finding fairly well (and note that it is only at these steps in our foreground bias estimation that the noise fields are used), it does lack larger scale noise correlations due to the atmosphere that may be important e.g. when it comes to detecting large, low redshift clusters\footnote{there has been extensive recent work on realistic noise simulations for single-array maps, by \citet{dr6-noise}, however we did not have access to realistic noise simulations for the deep coadds used for source and cluster detection during this work. Hence we perform the sensitivity test outlined in \app{app:noise_test}.}. We demonstrate in \app{app:noise_test} that our bias results are insensitive to using a noise model that includes large scale correlations.

We note here also that while cluster and point source model subtraction should mitigate the tSZ and CIB respectively, we do not implement a specific mitigation scheme for the kinetic Sunyaev-Zeldovich effect. However, it is included in our simulations, so will contribute to the foreground bias estimates presented here, and it is not expected to generate a significant bias at ACT DR6 accuracy levels \citep{ferraro18,cai22}.

\subsection{Adding \planck\ high frequency data and harmonic ILC}\label{sec:planckilc}

In addition to  our baseline approach, which uses only ACT maps to estimate the lensing (specifically the f090 and f150  data), we also investigate a frequency cleaning approach where we include high frequency data from \planck\ (353 and 545~GHz), where the CIB is brighter. To reduce the information on the primary CMB coming from \planck\ (i.e. to make this a lensing measurement that is mostly independent of \planck\ CMB information), we do not include lower frequency \planck\ data. 

To simulate the \planck\ data, we generate the total foregrounds-only maps at the \planck\ frequencies, and apply a point-source flux threshold of 304 and 555 mJy to the 353 and 545~GHz channels respectively (taking these thresholds from Table 1 of  \citealt{ade16}). We apply the ACT DR6 analysis mask to isolate the same sky area for the ACT and \planck\ data. Note that for the S10 simulations, no 545~GHz data was generated, therefore we simply scale the 353 GHz CIB simulation using an SED \citep{madhavacheril20} defined as 
\begin{equation}
    f_{\mathrm{CIB}}(\nu) \propto 
    \frac{\nu^{3+\beta}}{\exp{h\nu/k_B \Tcib-1}}
    \left(\left.\dfrac{\dx{B}(\nu,T)}{\dx{T}}\right\vert_{T_{CMB}}\right)^{-1}
    \label{eq:cib_spec}
\end{equation}
with $\beta = 1.2$, $\Tcib=24K$ and $B(\nu,T)$ is the Planck function. Note that since the CIB does not exactly follow such an $l$-independent spectrum, the use of this rescaling could underestimate foreground biases in tests of CIB-deprojection on the \sehgal\ simulations, however, the \websky\ simulations do have a more realistic 545~GHz channel. As for the other foreground components, the tSZ can be accurately scaled to 545~GHz with SED
\begin{equation} 
    f_{\mathrm{tSZ}}  \propto 
    \frac{h \nu}{k_B \Tcmb} 
    \frac{e^{h\nu/k_B \Tcmb}+1}{e^{h\nu/k_B \Tcmb}-1} - 4,
    \label{eq:tsz_spec}
\end{equation}
kSZ is frequency-independent in CMB units, and radio point-sources can be assumed to have negligble flux contribution at 545~GHz \citep{dunkley13}.

We produce a harmonic space \emph{internal  linear combination} (ILC, see e.g.  \citealt{remazeilles11,madhavacheril20}) of the maps that minimizes the variance of the output maps and additionally deprojects one or more foreground components by assuming an $l$-independent SED for those foreground components\footnote{A more advanced ILC analysis of the ACT DR6 maps, using needlet methods, is also underway and will be reported in \citet{dr6-ilc}}
. We use the same point-source and tSZ cluster-subtracted ACT maps as in our baseline analysis, while the \planck\ maps are only point-source subtracted. We investigate below the deprojection of the CIB and/or the tSZ, with SEDs given by \eqn{eq:cib_spec} and \eqn{eq:tsz_spec}, respectively. Again, we note that the assumption of an $l$-independent SED is only approximate in the case of the CIB. Minimizing the variance requires providing total power and cross-spectra for all the input maps; we construct these for frequencies $i$ and $j$ as
\begin{equation}
C_{l,ij}^{\mathrm{total}} = C_l^{\mathrm{cmb}} + C_{l,ij}^{\mathrm{fg}} + \delta_{ij} N_{l,\mathrm{i}}
\end{equation}
where $C_{l,ij}^{\mathrm{fg}}$ is the foreground (cross-)spectrum estimated from the simulation foreground-only maps, and $N_{l,\mathrm{i}}$ is the noise-only power spectrum, estimated from ACT DR6 noise simulations and the \planck\ noise-only simulations\footnote{available on NERSC at \url{/global/cfs/cdirs/cmb/data/planck2020}} provided with the NPIPE maps \citep{akrami20}.
Note that since we do not use signal power spectra measured directly from simulated maps containing lensed CMB signal, we will not incur any \emph{ILC-bias} \citep{delabrouille09}, which comes from the down-weighting of modes with high signal variance that can occur in an ILC. When generating ILC maps on the real DR6 data, we do use power spectra measured directly from the data, but apply aggressive smoothing to avoid this ILC bias.\footnote{See \citet{dr6-ilc} for further investigation of and mitigation methods for ILC bias.} 

\section{Simulation results: foreground biases to the lensing power spectrum}\label{sec:simpredictions}

Below we present our simulation-derived predicted biases due to extragalactic foregrounds, $\Delta \Clhatphi{\vecL}$. We calculate this following the methodology of \citet{schaan19}, who outline how each of the terms: primary bispectrum, secondary bispectrum and trispectrum can be estimated from simulations without noise due to the CMB or instrument noise realisation. The total bias is then simply the sum of these terms (see \eqn{eq:fg_expansion}).  

We estimate foreground biases for the temperature-only case as well as for our MV estimator, which is the power spectrum of $\hat{\phi}^{\mathrm{MV}}(\vecL)$, a lensing reconstruction which uses both temperature and polarisation data, given by a weighted sum of the reconstructions from all pairs $AB \in \left[TT,TE,TB,EE,EB\right]$:
\begin{equation}
    \hat{\phi}^{\mathrm{MV}}(\vecL) = \sum_{AB} w_{AB} \hat{\phi}^{AB}(\vecL)
    \label{eq:phimv}
\end{equation}
(see \citet{dr6-lensing-auto} for the exact details of these weights in the bias-hardened case).


We weight each $\hat{\phi}^{AB}(\vecL)$ by the inverse of the normalization (which is equivalent to its response to lensing). The bias on the MV lensing power spectrum estimate is then \citep{darwish21}
\begin{align}
\begin{split}
    \Delta \CLhatphiX{MV} &= 
    \left[ (w_L^{TT})^2 \Delta \CLhatphiX{TTTT} \right.\\
 &+ \frac{1}{2}\sum_{XY \in (EE, BB, TE)} w_L^{TT} w_L^{XY} P^{TT}_L\\
 &+ \left. w_L^{TT}w_L^{TE} S^{TTTE}_L + w_L^{TE}w_L^{TE} S^{TETE}_L  \right.\\
 &+ \left. w_L^{TT}w_L^{TB} S^{TTTB}_L + w_L^{TB}w_L^{TB} S^{TBTB}_L  \right.\\
  &+ \left. w_L^{TE}w_L^{TB} S^{TETB}_L + w_L^{TB}w_L^{TB} S^{TBTB}_L
 \right].
\end{split}
\label{eq:mvbias}
\end{align}
where $\Delta \CLhatphiX{TTTT}$ is the bias on the temperature-only estimator. $P^{TT}_L$ is the primary bispectrum bias on the temperature-only estimator, and is the dominant additional bias term for the MV estimator, coming from the correlation of foreground temperature with the lensing convergence reconstructed from polarisation, $\left< Q(\Tfg, \Tfg) \hat{\kappa}^{\mathrm{pol}} \right>$. There are additional secondary bispectrum terms, $S^{ABCD}$ of the form
\begin{equation}
    \left< Q(A^{\mathrm{CMB}}, B^{\mathrm{fg}}) Q(C^{\mathrm{CMB}}, D^{\mathrm{fg}}) \right>
\end{equation}
with $A,B,C,D \in [T,E,B]$. Even the largest of these, $S^{TTTE}$ is negligible for our analysis.

We also show the bias, $\Delta\Alens$ in the inferred lensing power spectrum amplitude $\Alens$, which we approximate as 
\begin{equation}
    \Delta\Alens = \frac{\sum_L \sigma^{-2}_L \Delta \Clhatphi{L}\Clphi{L}}{\sum_L  (\Clphi{L}/\sigma_L)^2}
\end{equation}
for the case of a diagonal covariance matrix on $\Clhatphi{L}$ with diagonal elements $\sigma_L$, and true signal $\Clphi{L}$.
The uncertainty on $\Alens$ is given by $\left(\sum_L  (\Clphi{L}/\sigma_L)^2\right)^{-0.5}$, so 
\begin{equation}
    \frac{\Delta\Alens}{\sigma(\Alens)} =
    \frac{\sum_L \sigma^{-2}_L \Delta \Clhatphi{L}\Clphi{L}}{\sqrt{\sum_L  (\Clphi{L}/\sigma_L)^2}}.
\end{equation}
In this work, we calculate $\sigma(\Alens)$ based on an approximate covariance (assuming a reconstruction noise based on the analytic full-sky $N^0$, scaled appropriately for the ACT sky fraction); this allows us to quickly re-compute the covariance for the various different mitigation strategies explored here. This results in an underestimation of $\sigma(\Alens)$ by $\sim 10\%$ for the temperature-only case and $\sim 5\%$ for the MV case, relative to the more accurate uncertainty for the DR6 data, recovered using simulations  in \citet{dr6-lensing-auto}, which we will call $\sigma^{\mathrm{data}}(\Alens)$. Given computational limitations we do not have $\sigma^{\mathrm{data}}(\Alens)$ for most of the analysis variations tested here. Therefore, when we report $\Alens$ uncertainties in \tab{tab:Alensbias}, rather than quoting these underestimated uncertainties, we scale them by the ratio $\sigma^{\mathrm{data}}(\Alens) / \sigma(\Alens)$ for the baseline analysis, for which we have both these approximate uncertainties and the accurate simulation-based estimates.

\begin{figure*}
    \centering
    \includegraphics[width=0.98\columnwidth]{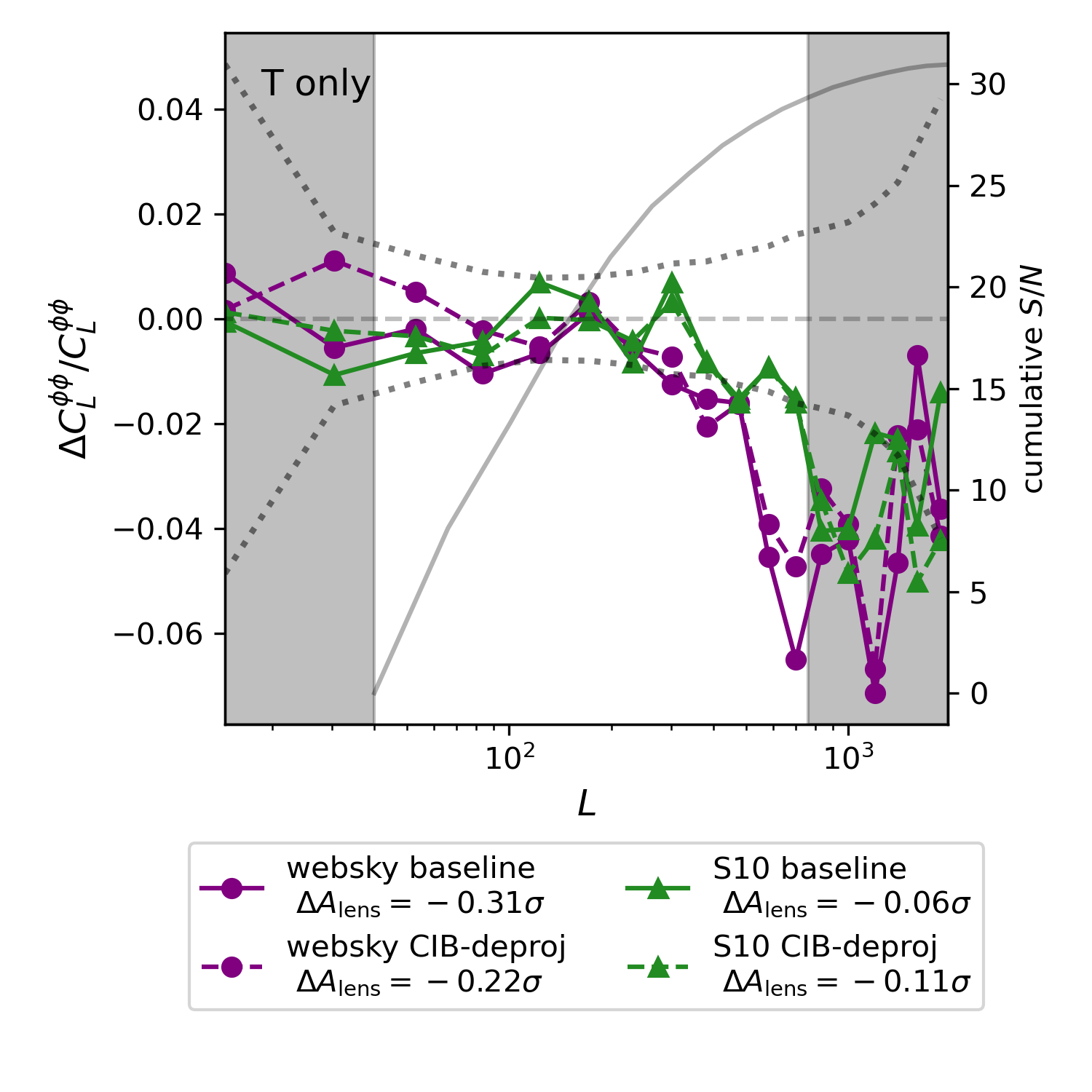}
    \includegraphics[width=0.98\columnwidth]{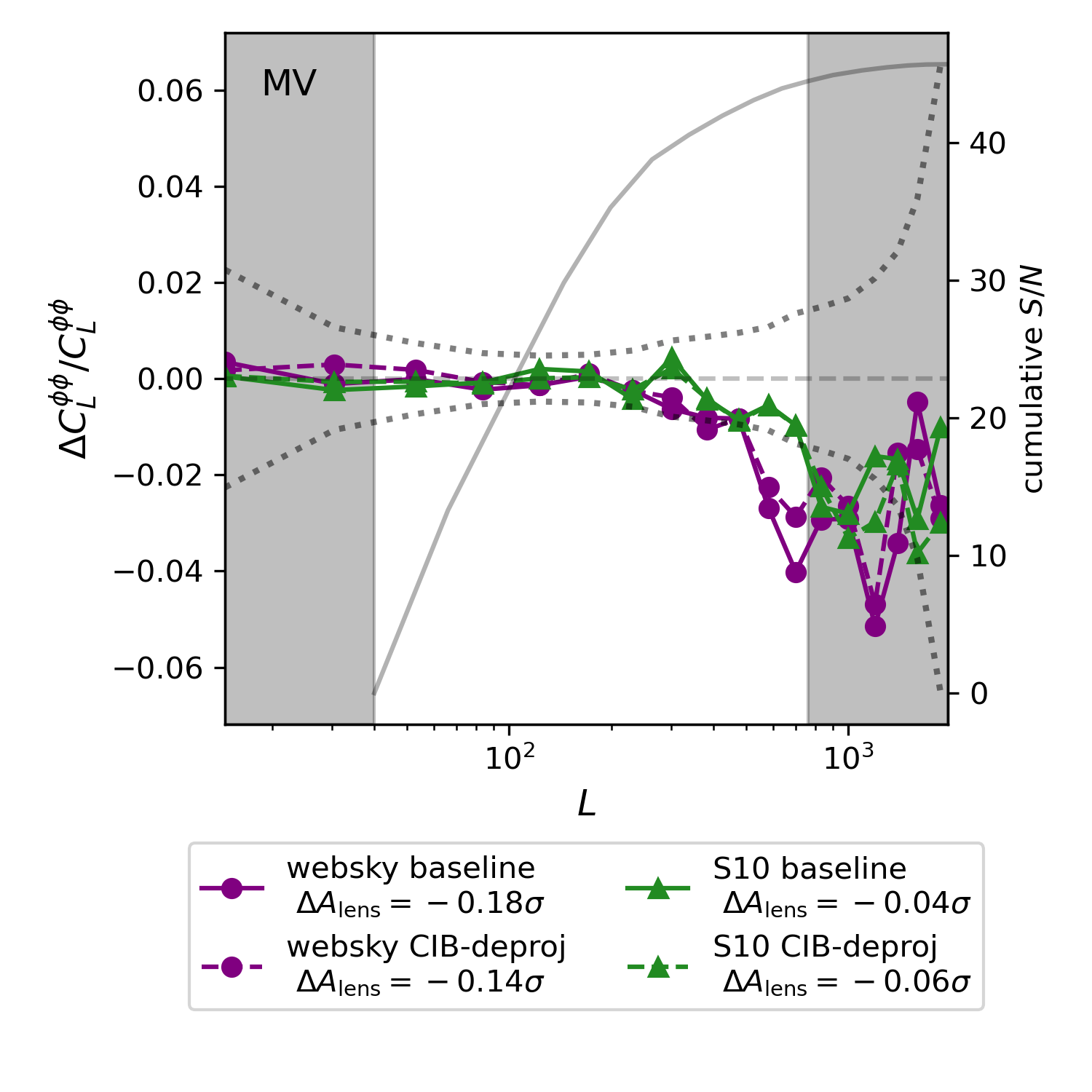}
    \caption{Fractional bias due to extragalactic foregrounds to the estimated CMB lensing power spectrum, for an ACT DR6-like analysis. Left panel: The bias for the temperature-only power spectrum. Right panel: The bias for the MV power spectrum (see \eqn{eq:mvbias}), which is the measurement used for cosmological inference in \citet{dr6-lensing-auto,dr6-lensing-cosmo}. In both panels, purple circles with solid connecting lines indicated biases estimated from the \websky\ \citep{websky} simulations, green triangles with solid connecting lines indicated biases estimated from the \sehgal\ \citep{Sehgal_2010} simulations. Dashed lines indicate the foreground biases for the CIB-deprojected analysis variation described in \sect{sec:cibdeproj}.
     The grey dotted line indicates the $1\sigma$ uncertainty of the DR6 bandpower measurement divided by 10, and the solid grey line indicates the cumulative $S/N$ when only scales up to $L$ are included. The grey shaded regions indicate scales not used in the cosmological inference, as described in \citet{dr6-lensing-auto}. For all cases,  we quote in the legend the total bias in inferred lensing power spectrum amplitude, $\Alens$ in units of the DR6 $1\sigma$ uncertainty on that quantity. Dashed grey lines indicate zero bias.}
    \label{fig:baseline1}
\end{figure*}

\subsection{Baseline Analysis}

In the baseline power spectrum analysis of \citet{dr6-lensing-auto} and \citet{dr6-lensing-cosmo}, we perform lensing reconstruction on a weighted coadd of the f090 and f150  maps, with each frequency $i$ weighted by the inverse of its 1-dimensional harmonic-space noise power spectrum, $N_l^i$ as estimated from simulations. The inverse noise filter used in the quadratic estimator is diagonal in harmonic space, and uses the total power spectrum $C_l^{\mathrm{tot}} = C_l^{\mathrm{cmb}}+N_l^{\mathrm{coadd}}$, 
the sum of a fiducial CMB power spectrum and the coadd noise power spectrum. 

We perform these same steps on the simulated foreground-only maps used to estimate foreground biases here i.e. we use the same $l$-dependent weights to coadd the f090 and f150 maps, and the same filtering of the maps entering the quadratic estimator (rather than using the total power spectrum of the foreground-only maps themselves, for example, since the aim here is to use the same weighting of modes as used in the real data reconstruction).

The solid lines in \fig{fig:baseline1} show the predicted bias to the lensing reconstruction power spectrum from temperature data only (left panel), and for the MV estimator (right panel), as a fraction of the expected $C_L^{\kappa\kappa}$ signal, for the baseline method. For both the \websky\  (purple lines) and S10 (green lines) simulations, the absolute fractional biases are within 2\% up to $L\approx500$, which is where most of the \snr\ of the DR6 measurement will come from (the solid, light grey line indicates the cumulative \snr\ as a function of the maximum $L$ included). For \websky\, the fractional bias does start to exceed that level at higher $L$, and supports our pre-unblinding decision to limit the baseline analysis to $L<=763$. For guidance, the dotted grey line indicates the DR6 $1\sigma$ uncertainty divided by 10, indicating that biases for each bandpower are mostly below $0.1\sigma$, except for \websky\ at $L>500$, where they are still well below $0.5\sigma$. 

We emphasize that the result for the MV estimator is most important, since that is what we use for cosmological inference in \citet{dr6-lensing-auto,dr6-lensing-cosmo}. However, it is encouraging that biases in the temperature-only estimator are also small, since it allows us to use the consistency of the MV and temperature-only measurements as a test of other systematics that affect temperature and polarisation differently. 

\begin{figure}
\centering
 \includegraphics[width=0.9\columnwidth]{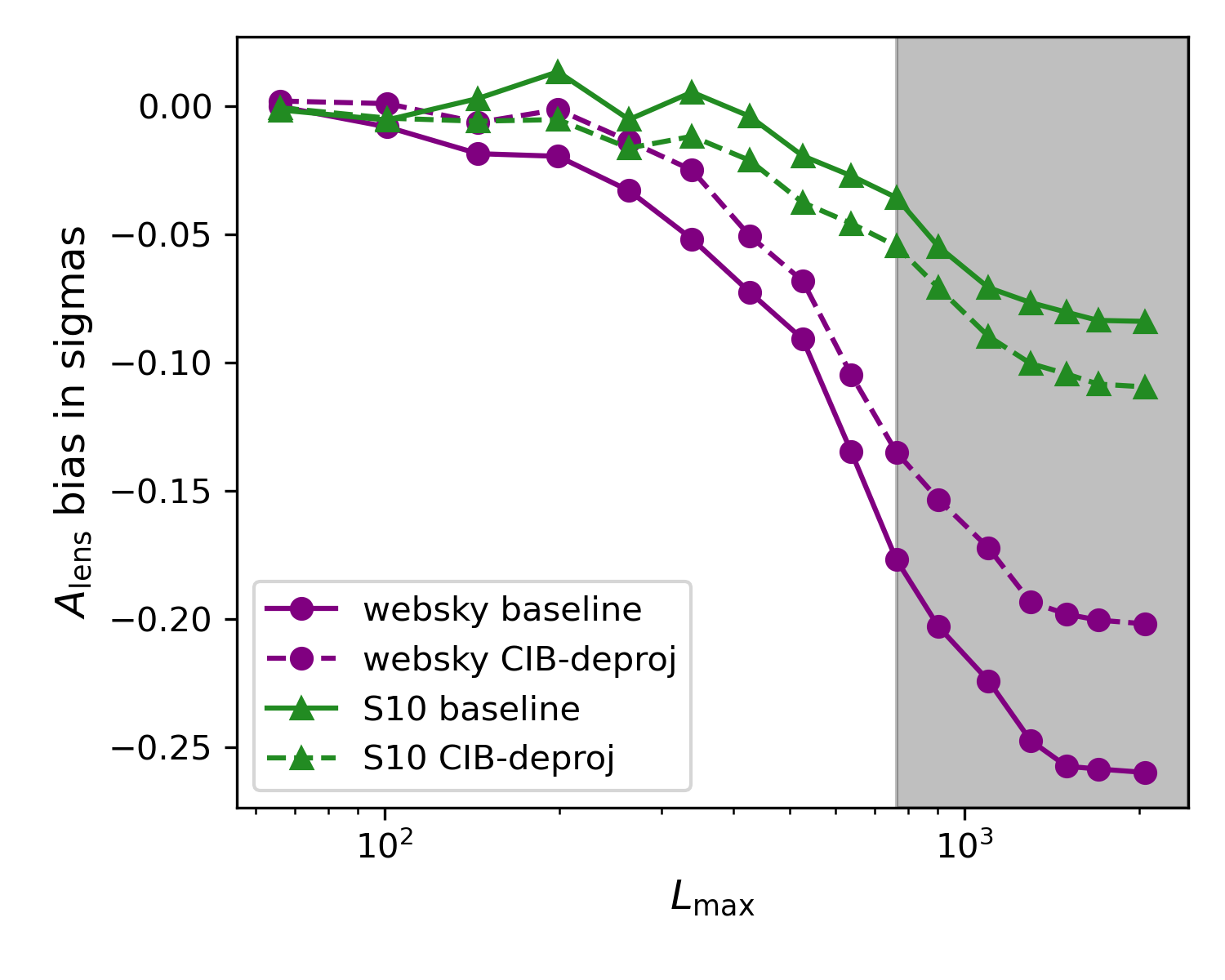}
    \caption{The key result of our simulation tests - the bias in inferred lensing power spectrum, $\Alens$, in units of the $1\sigma$ uncertainty, as a function of the maximum scale, $\Lmax$ (for the MV estimator). Purple circles and solid (dashed) lines show the prediction from \websky\ for the baseline (CIB-deprojected analysis), while green triangles show the predictions for the \sehgal\ simulations. The grey shaded regions show ranges not included in the baseline cosmology analysis in \citet{dr6-lensing-auto,dr6-lensing-cosmo}}
    \label{fig:simbias_highL}
\end{figure}

To more concisely quantify the cumulative impact of these scale-dependent fractional biases, \fig{fig:simbias_highL} shows the bias in inferred $\Alens$ as a function of the maximum $L$ used, $\Lmax$. For our baseline range of $\Lmax=763$, we  report in \tab{tab:Alensbias} and the legend to  \fig{fig:baseline1} the bias in inferred $A_{\mathrm{lens}}$, which is well below $1\sigma$ for both simulations, for both the temperature-only (-0.31$\sigma$ for \websky\ and -0.06$\sigma$ for \sehgal) and MV cases (-0.18$\sigma$ for \websky\ and -0.04$\sigma$ for \sehgal).

When including higher-$L$ scales, the predicted absolute biases are still modest, not exceeding $0.3\sigma$ when including up to $\Lmax=1300$. We therefore believe it is reasonable for the lensing power spectrum analyses in \citet{dr6-lensing-auto,dr6-lensing-cosmo} to also consider the extended range, $\Lmax=1300$.

\begin{figure*}
\centering
 \includegraphics[width=0.95\textwidth]{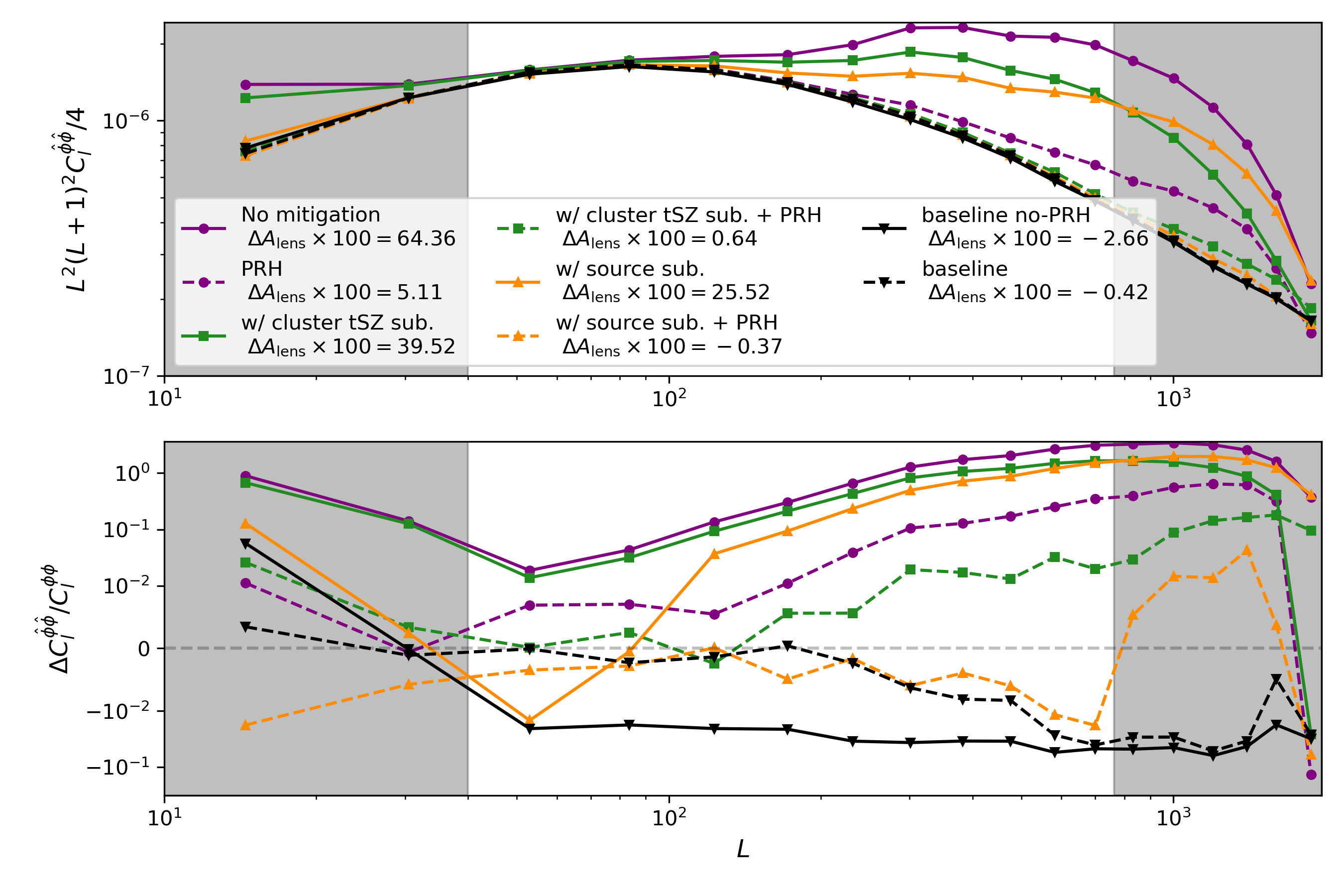}
    \caption{A demonstration of our three main mitigation strategies at work: the total recovered lensing power spectrum (top panel), and its fractional bias with respect to the truth (bottom panel), for the temperature-only case, as predicted from the \websky\ simulation. Purple lines labelled ``no mitigation'' show the case where no foreground mitigation is applied. Green (orange) lines have clusters (point source) models subtracted while black solid lines have both subtracted. Dashed lines indicated the use of a profile-hardened lensing estimator (labelled "PRH", see Sections \ref{sec:bh} and \ref{sec:actstuff} for details). In the legend we also note the fractional bias in lensing power spectrum amplitude, $\Delta\Alens$, in percent. Dashed grey lines indicate zero bias. The grey shaded regions indicate scales not used in the cosmological inference, as described in \citet{dr6-lensing-auto}.}
    \label{fig:mitigation_demo}
\end{figure*}

To provide insight into the impact of each of our mitigation strategies, \fig{fig:mitigation_demo} shows the foreground biases predicted from \websky\ when no mitigation is used (purple solid lines), when cluster model subtraction is included (green solid lines), when source subtraction is included (orange solid lines), and when both cluster models and sources are subtracted (black solid lines). For each of these cases we also show as dashed lines the case where the profile-hardened estimator is used, as in our baseline analysis. Without profile hardening, both cluster and model subtraction are required to get the biases down to the few percent level. Using profile hardening in addition enables us to reduces foreground bias of $\Alens$ to below $1\%$.


\begin{table*}
\centering
\begin{tabular}{|l |l |c |c| c | c |}
\hline
Simulation & Analysis version & $\Delta\Alens^{TT}$x100 & $\sigma^{TT}(\Alens)$x100 & $\Delta\Alens^{MV}$x100 & $\sigma^{MV}(\Alens)$x100\\
\hline
websky & baseline & -1.2 & 3.7 & -0.42 & 2.3\\
websky & CIB-deproj incl. \textit{Planck} & -0.83 & 3.8 & -0.32 & 2.3\\
S10 & baseline & -0.24 & 3.7 & -0.088 & 2.3\\
S10 & CIB-deproj incl. \textit{Planck} & -0.41 & 3.8 & -0.13 & 2.3\\
\hline
websky & tSZ-deproj + PSH & -1.9 & 32 & -0.12 & 4.2\\
websky & D21-tSZ-deproj + PSH & 1.4 & 7.4 & 0.56 & 2.3\\
S10 & tSZ-deproj + PSH & 0.19 & 25 & -0.024 & 4.1\\
S10 & D21-tSZ-deproj + PSH & 1.2 & 6.7 & 0.45 & 2.3\\
\hline
websky & CIB-deproj incl. \textit{Planck} + PRH & -0.81 & 3.8 & -0.31 & 2.3\\
websky & D21-CIB-deproj incl. \textit{Planck} + PRH & 0.61 & 3.7 & 0.2 & 2.3\\
S10 & CIB-deproj incl. \textit{Planck} + PRH & -0.49 & 3.8 & -0.17 & 2.3\\
S10 & D21-CIB-deproj incl. \textit{Planck} + PRH & 0.54 & 3.7 & 0.18 & 2.3\\
\hline
websky & tSZ and CIB-deproj incl. \textit{Planck} + PRH & 8.7 & 12 & 0.53 & 3.6\\
websky & D21 tSZ and CIB-deproj incl. \textit{Planck} + PRH & 0.63 & 6.2 & 0.22 & 2.3\\
S10 & tSZ and CIB-deproj incl. \textit{Planck} + PRH & 1.5 & 11 & 0.098 & 3.5\\
S10 & D21 tSZ and CIB-deproj incl. \textit{Planck} + PRH & 0.24 & 5.9 & 0.09 & 2.3\\
 \hline
\end{tabular}
\caption{The bias and uncertainty on the inferred lensing power spectrum amplitude, $\Alens$, predicted from the \websky\ and \sehgal\ simulations, for the TT and MV estimators, for a range of analysis variations. Note in the unbiased case, $\Alens=1$, so $\Delta\Alens$ constitutes a fractional bias. ``PSH'' and ``PRH'' indicated point-source hardeneing and profile hardening respectively (see \sect{sec:bh} for details).}
\label{tab:Alensbias}
\end{table*}

\subsection{CIB-deprojected Analysis}\label{sec:cibdeproj}

In addition to the baseline analysis, we perform the lensing reconstruction on CIB-deprojected maps, using the methodology described in \sect{sec:planckilc}. We include the \planck\ 353 GHz and 545~GHz channels, in which the CIB has much higher amplitude than at ACT frequencies. In approximate terms, these high frequency channels provide maps of the CIB that are ``subtracted" out by the CIB deprojection, while providing very little information on the CMB, Thus this analysis is still largely   independent of \planck\ CMB information (it is to maintain this that we do not use the \planck\ 217 GHz channel). Note that for the ACT maps, point sources and tSZ clusters are still modelled and subtracted, as in our baseline approach, while for the Planck data only point-sources are subtracted. We still also use the same bias-hardened quadratic estimator as in our baseline approach.

The dashed lines in \fig{fig:baseline1} show the fractional biases to $\Clhatkappa{L}$ predicted from the \websky\ (purple dashed) and \sehgal\ (green dashed) simulations. For both simulations the predicted biases are quite similar to the baseline case,  suggesting that this CIB-deprojection approach is also a useful option to use on the DR6 data.

\subsection{Other options}\label{sec:other_options}
We report in \tab{tab:Alensbias} statistics on the $\Alens$ bias for various other mitigation strategies, summarized below:

\begin{itemize}
\item{tSZ-deprojection with only the two ACT channels used here (f090 and f150) greatly increases the reconstruction noise (by a factor of $\sim10$ for the temperature only estimator). This is because the tSZ amplitude at 150 GHz is roughly half that at 90 GHz, therefore to null the tSZ requires weighting the  noisier 150 GHz with roughly twice the weight as the 90 GHz data. In addition, tSZ-deprojection upweights the CIB (which is stronger at 150GHz), resulting in $\Alens$ biases at the percent level. For these reasons, simply using tSZ-deprojected maps, with only ACT channels, in both legs of the quadratic estimator is not a viable option. The noise cost is reduced for the D21 estimator, but is still a factor of $\sim 2$ for the temperature estimator. The biases are slightly smaller for these tSZ-deprojected cases when performing point-source hardening (indicated in \tab{tab:Alensbias} by `PSH') rather than profile hardening (indicated in \tab{tab:Alensbias} by `PRH'), presumably since it is better suited to the dusty galaxies responsible for the CIB. When performing tSZ-deprojection, and not also deprojecting the CIB, including high frequency data from \planck\ is not useful, because  the CIB contamination from these high frequencies becomes very large. Hence we do not show results for that option here.}
\item{
CIB-deprojection using only ACT channels similarly has a large noise cost; including \planck\ channels solves this by effectively providing a relatively high signal-to-noise CIB map to subtract (as described in \sect{sec:cibdeproj}). As well as our CIB-deprojection option, where both temperature maps in the quadratic estimator are cleaned, we apply the D21 estimator for this case. For this case, labelled ``websky/S10 D21-CIB-deproj incl. \planck\ + PRH'',  percent-level $\Alens$ biases remain. On inspecting the contributions to this bias, we find this is due to an increased trispectrum term relative to the fully-cleaned estimator (the primary and secondary terms are approximately unchanged). Since we would not expect the tSZ or CIB trispectra to increase for the D21 estimator relative to the fully CIB-cleaned estimator, this is likely due to the presence of terms of the form $\left<Q(T^{\mathrm{CIB}}T^{\mathrm{tSZ}})Q(T^{\mathrm{CIB}}T^{\mathrm{tSZ}})\right>$ for the D21 estimator (this term is approximately zero  for the usual CIB-deprojected estimator where all legs have an approximate CIB spectrum deprojected).
}
\item{When including the \planck\ high frequency data, we have sufficient degrees of freedom to deproject both tSZ and CIB. However, there is a large noise cost to doing so, resulting in an $\Alens$ uncertainty of $\sim 5$ times larger than the baseline analysis (and $\sim 2$ times larger for the D21 estimator) for the temperature-only estimator.}
\end{itemize}

\section{Data nulls}\label{sec:nulls}

Having settled on our mitigation strategies based on the predicted biases from \websky\ and S10 simulations (\sect{sec:simpredictions}), we now turn to the DR6 data to further validate the performance  of these strategies. In \sect{sec:freqdiffnull} we present null tests involving differences of ACT single-frequency information (in which the CMB lensing signal is nulled), then in \sect{sec:cibnull} we compare the DR6 bandpowers estimated from a CIB-deprojected map to our baseline DR6 result.

\subsection{Frequency difference tests}\label{sec:freqdiffnull}

The two ACT channels we have used have different sensitivities to foregrounds, in particular, the tSZ has a higher amplitude at 90 GHz than at 150 GHz, and the opposite is true for the CIB. We can  use differences between the single-frequency data to form null tests, since the lensed CMB signal is nulled in these differences, while foregrounds are not.
If our mitigation strategies are working well, however, we will still find that our lensing estimators applied to these foreground-only maps return null signals. We consider three such null tests in the following that have  somewhat different sensitivity to the different terms in the foreground expansion of \eqn{eq:fg_expansion}.

\subsubsection{Null map auto spectrum}\label{sec:nullmap}

\begin{figure}
    \centering
    \includegraphics[width=0.95\columnwidth]{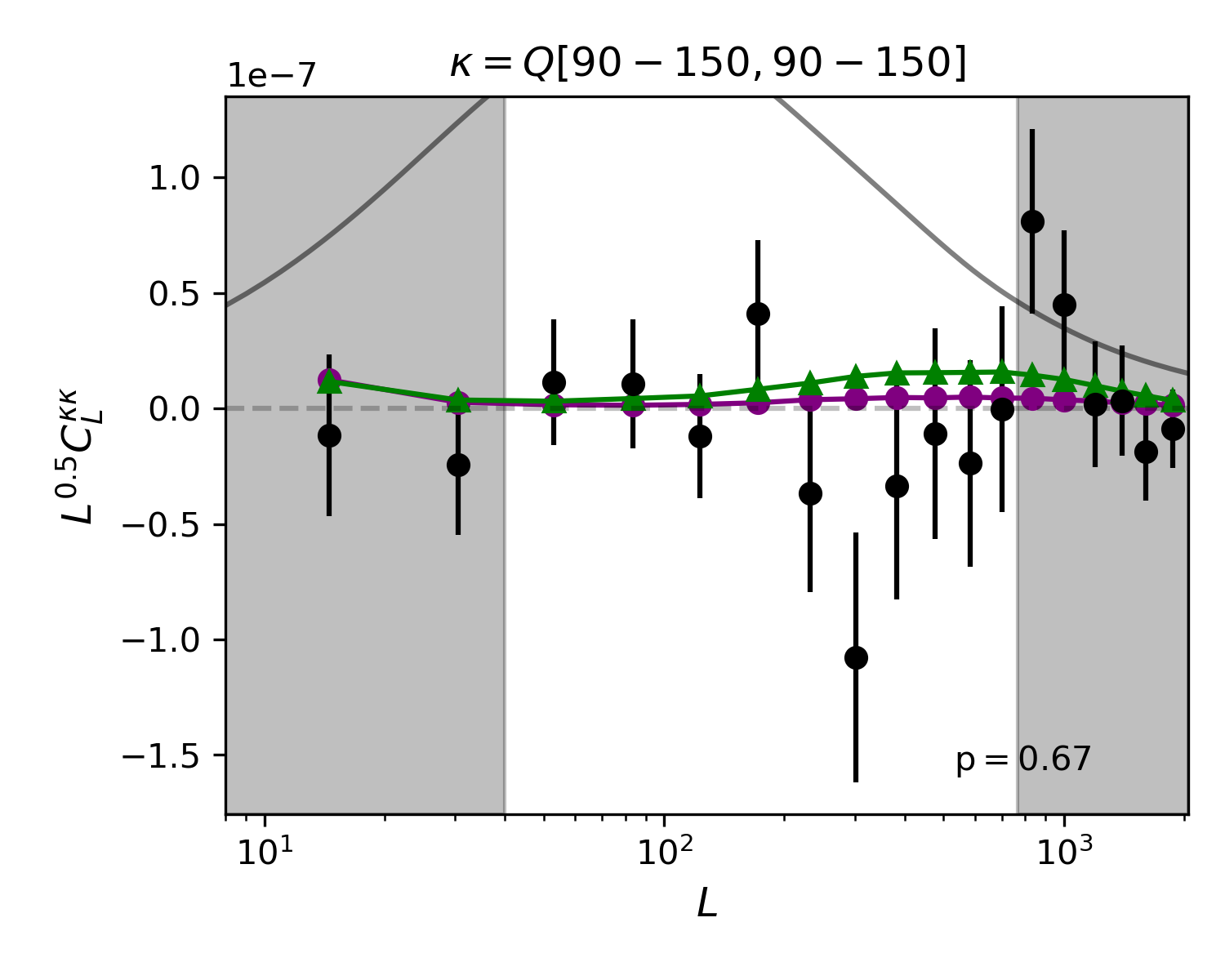}
    \includegraphics[width=0.95\columnwidth]{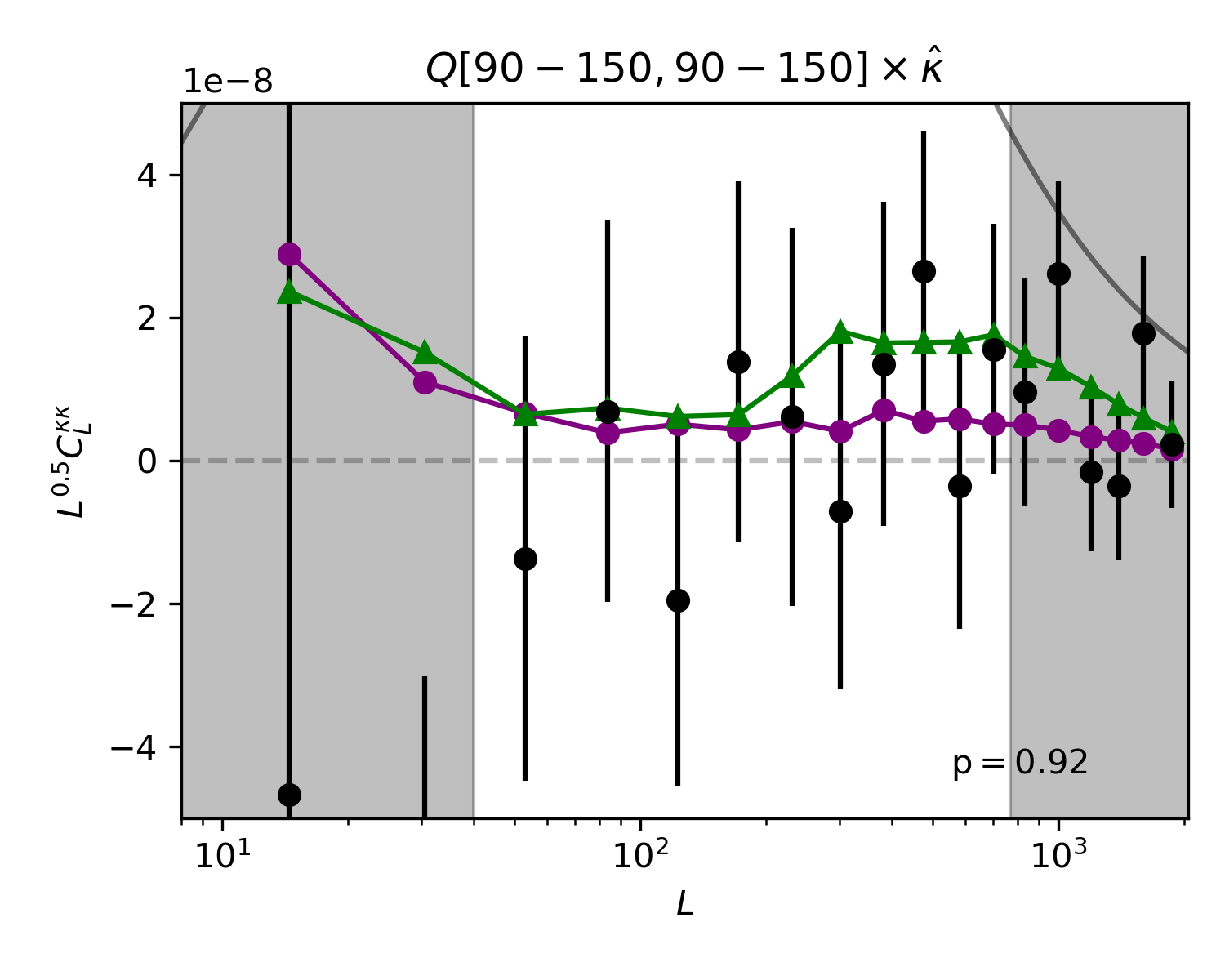}
    \includegraphics[width=0.95\columnwidth]{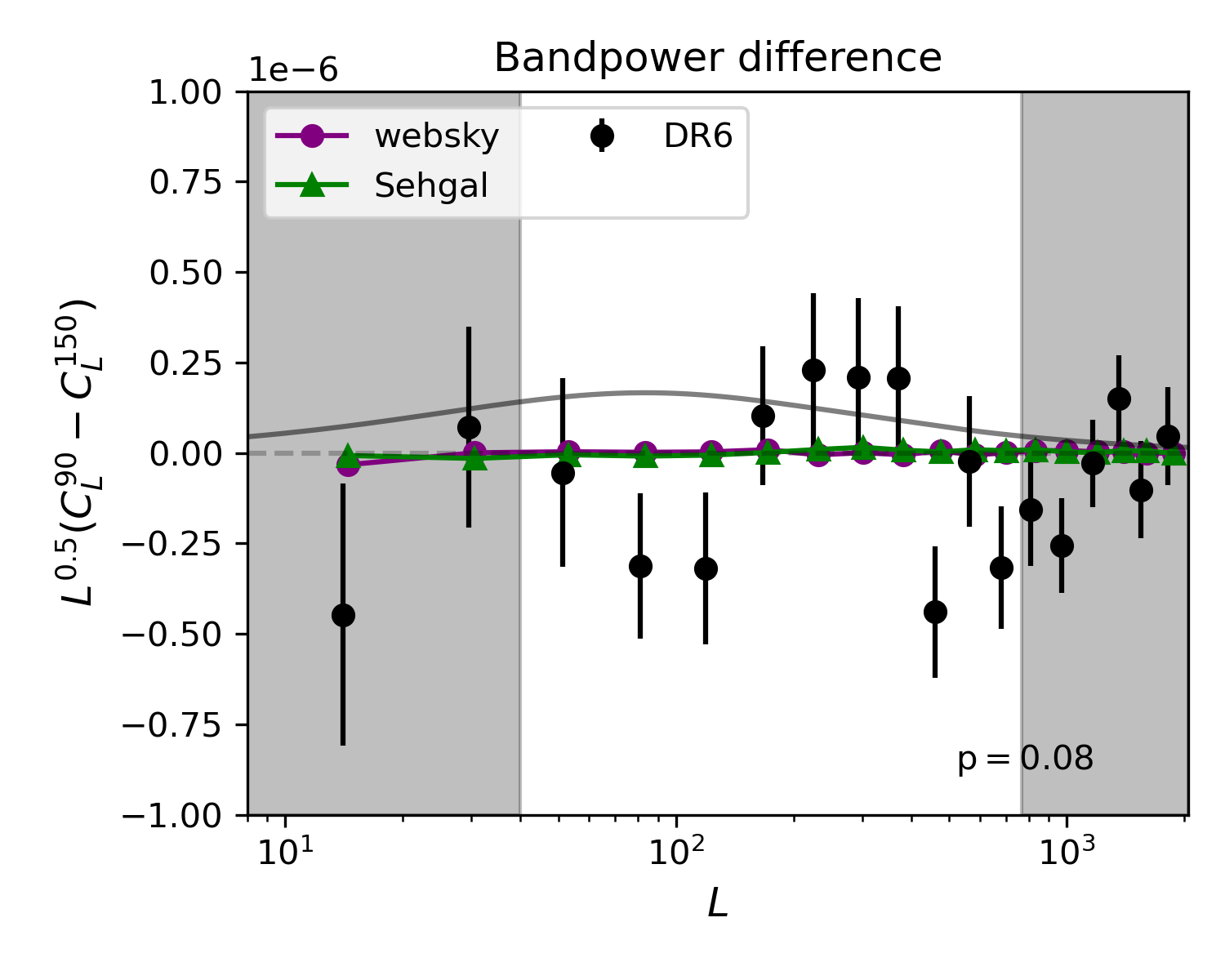}
    \caption{Frequency-difference null tests, designed to test the sensitivity of our lensing power spectrum estimation to differences in the foreground contributions to the f090 and f150 temperature data. In each panel, the CMB lensing signal is nulled in a different way,  see sections \ref{sec:nullmap}-\ref{sec:nullbp} for details. All the null tests perform satisfactorily i.e. the p-values are $\geq0.05$. In each panel, the DR6 data are shown as black circles, and the simulation predictions for \websky\ and \sehgal\ are shown as purple circles and green triangles respectively.  Solid grey lines indicate the theoretical lensing power spectrum multiplied by 0.1, and dashed grey lines indicate zero bias. The grey shaded regions indicate scales not used in the cosmological inference, as described in \citet{dr6-lensing-auto}.}
    \label{fig:freqdiffnulls}
\end{figure}

We perform reconstruction on the difference of the individual frequency maps (using temperature data only to maximise sensitivity to extragalactic foregrounds), and take the power spectrum:
\begin{equation}
\begin{split}
    C_L^{\mathrm{null},1} = \langle & Q(T^{90}-T^{150},T^{90}-T^{150})  \\ 
     \times&Q(T^{90}-T^{150},T^{90}-T^{150})  \rangle .
\end{split}
\end{equation}
Since CMB signal is nulled in the input maps to this reconstruction, this measurement is insensitive to the bispectrum terms and depends only on the trispectrum of the frequency difference map $T^{90}-T^{150}=T^{\mathrm{fg},90}-T^{\mathrm{fg},150}$ where $T^{\mathrm{fg},i}$ is the foreground contribution to frequency $i$. The top panel of \fig{fig:freqdiffnulls} shows this measurement on DR6 data, showing a null signal, as well as the predictions from the \websky\ and S10 simulations. The solid grey line indicates the $C_L^{\kappa\kappa}$ theory prediction divided by 10, so any foreground trispectrum hiding beneath the noise here is well below the true lensing signal.
    
\subsubsection{Null map $\times$ $\hat{\kappa}^{MV}$ spectrum}\label{sec:nullxk}

We cross-correlated the reconstruction based on the frequency difference map with the baseline (i.e. non-nulled) reconstruction, $\hat{\kappa}^{MV}$:
\begin{equation}
    C_L^{\mathrm{null},2} =\left<Q(T^{90}-T^{150},T^{90}-T^{150})\hat{\kappa}^{MV}\right>.
\end{equation}
If $\hat{\kappa}^{MV}$ was the true $\kappa$,  this measurement would be sensitive only to the difference in the primary bispectra contributions for the two frequencies. In fact since $\hat{\kappa}^{MV}$ will have small foreground biases, there will also be a trispectrum contribution present of the form
\begin{equation}
\begin{split}
    \langle & Q( T^{\mathrm{fg},90}-T^{\mathrm{fg},150},T^{\mathrm{fg},90}-T^{\mathrm{fg},150} ) \\ 
     &Q( T^{\mathrm{fg},\mathrm{coadd}},T^{\mathrm{fg},\mathrm{coadd}})\rangle
\end{split}
\end{equation}
The middle panel of \fig{fig:freqdiffnulls} shows this measurement on DR6 data, as well as the predictions from the \websky\ and S10 simulations; showing a null signal. Again, the solid grey line indicates the $C_L^{\kappa\kappa}$ theory prediction divided by 10.

\subsubsection{Bandpower frequency difference}\label{sec:nullbp}
We take the difference of the auto-spectra of reconstructions performed on single-frequency maps i.e.
\begin{equation}
    C_L^{\mathrm{null},3} =C_L^{\hat{\kappa}\hat{\kappa},90\mathrm{GHz}} -C_L^{\hat{\kappa}\hat{\kappa},150\mathrm{GHz}}
\end{equation}
This null is sensitive to all three contributions (i.e. primary bispectrum, secondary bispectrum and trispectrum) in \eqn{eq:fg_expansion}, where one would substitute $\Tfg =T^{\mathrm{fg},90}-T^{\mathrm{fg},150}$ to model the result of this measurement. 

The bottom panel of \fig{fig:freqdiffnulls} shows this measurement on DR6 data, as well as the predictions from the \websky\ and S10 simulations; showing a null-signal. This test is  noisier than the first two, since lensed CMB is not nulled at the map level. We note that one could form other null measurements, for example those of the form 
\begin{equation}
    \left<Q(T^{90}-T^{150}, T^{\mathrm{coadd}})Q(T^{90}-T^{150}, T^{\mathrm{coadd}}) \right>
\end{equation}
in order to target and disentangle the secondary bispectrum contribution, but given we find this to be very small in simulations, we leave such exercises for future work.


\subsection{Consistency of CIB-deprojected analysis}\label{sec:cibnull}

As described in \sect{sec:simpredictions}, the CIB-deprojected version of our analysis performs well on simulations, with predicted systematic bias in the lensing amplitude well below our statistical uncertainties. This prediction of course depends  on the simulations, which may contain some inaccuracies in modeling extragalactic foreground components\footnote{Note that these simulations are typically tuned to real data at the power spectrum level, while the extragalactic foreground biases here depend on higher order statistics of the foreground fields.}. It is therefore useful to test for consistency of the CIB-deprojected and baseline analysis on the DR6 data. We generate CIB-deprojected temperature maps by combining the DR6 data with \planck\ 353~GHz and 545~GHz data, using the same procedure as described in \sect{sec:planckilc}. As explained in \citet{dr6-lensing-auto}, we additionally remove a small area at the edge of our mask that has strong features in the high frequency \planck\ data due to Galactic dust.  We generated 600 simulations of these maps, using the \planck\ \textsc{npipe} noise simulations provided by \citet{plancknpipe}; these are used for the $N^0$ and mean-field bias corrections (see \citealt{dr6-lensing-auto} for details).

The deprojected temperature maps  are then used in our lensing reconstruction (including profile hardening as in the baseline analysis) and lensing power spectrum estimation (in combination with the same polarisation data as is used in the baseline analysis). 

\begin{figure}
    \centering
    \includegraphics[width=0.9\columnwidth]{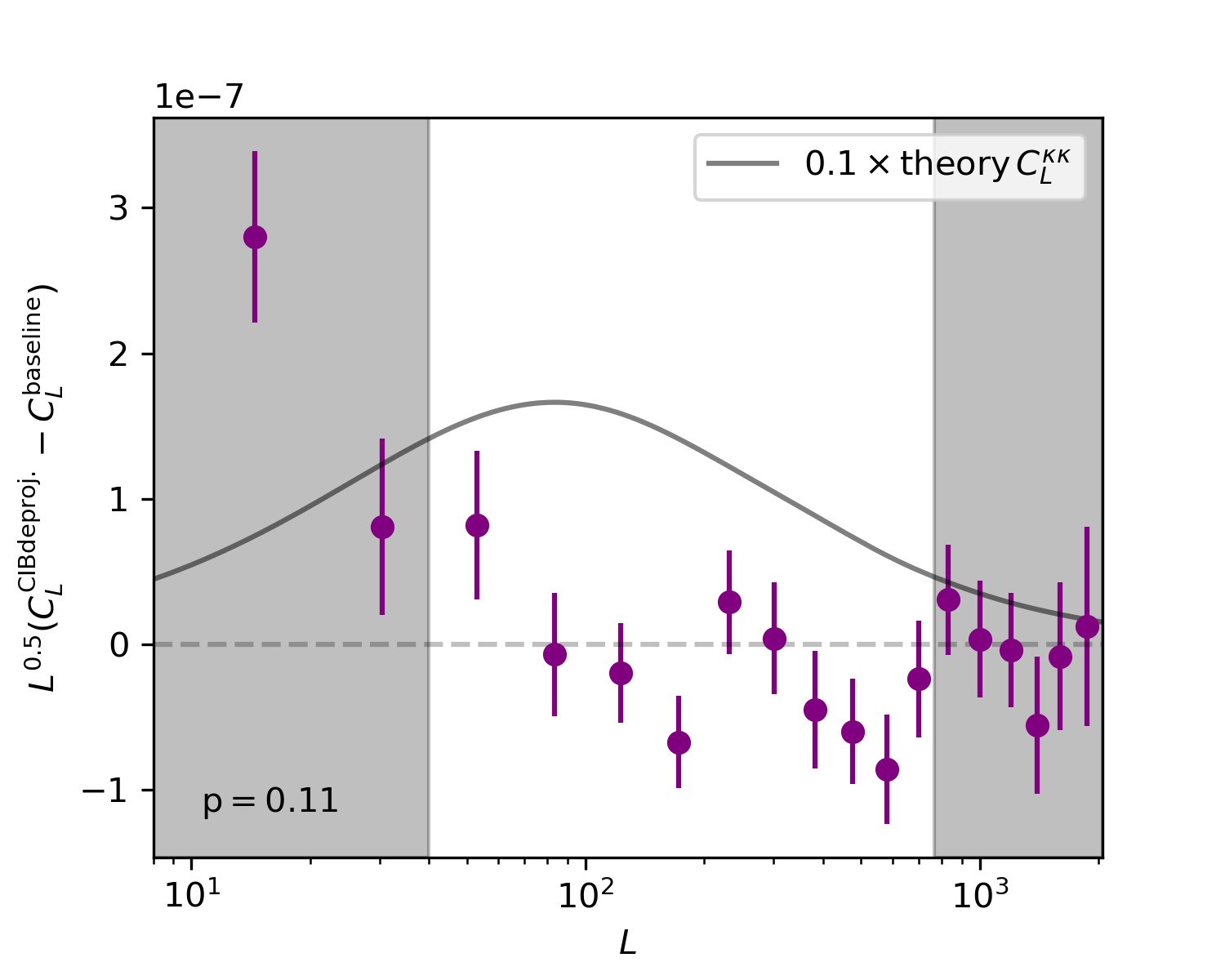}
    \caption{The difference in the ACT DR6 lensing power spectrum estimated from CIB-deprojected maps (as described in \sect{sec:cibnull}), and the baseline measurement. Grey shaded regions indicated $L$-ranges excluded from the baseline analysis. The $p$-value = 0.11 for the null result (including only data points within the baseline analysis range). The dashed grey line indicates zero bias, while the solid grey line is the prediced lensing power spectrum divided by 10. The grey shaded regions indicate scales not used in the cosmological inference, as described in \citet{dr6-lensing-auto}.}
    \label{fig:cibdeproj}
\end{figure}

\fig{fig:cibdeproj} compares this CIB-deprojected measurement to the baseline measurement (which does not perform frequency-cleaning), finding no evidence for inconsistency. This result implies that CIB contamination is unlikely to be significant in our baseline analysis.

We note that we do not perform an equivalent consistency test for tSZ-deprojection; as described in \sect{sec:other_options}, tSZ-deprojection incurs a very large noise cost when including only ACT data, and can generate very large biases due to boosting the CIB when including higher frequencies from \planck. While joint tSZ and CIB deprojection could address this large bias, we do not find this an effective strategy in our simulation tests since it incurs large noise costs and non-negligible  biases. 

\section{Conclusions}\label{sec:conclusion}

Extragalactic foregrounds are a potentially significant source of systematic bias in CMB lensing estimation, especially for temperature-dominated current datasets such as ACT. We have argued that the mitigation strategies implemented for the DR6 lensing power spectrum analysis (see \citealt{dr6-lensing-auto,dr6-lensing-cosmo}) ensure negligible bias due to extragalactic foregrounds. These mitigation strategies are: i) finding (using a matched-filter algorithm) and subtracting models for $S/N>4$ point-sources to remove contamination from radio sources and dusty galaxies (or the CIB), ii) finding (using a matched-filter algorithm) and subtracting models of galaxy clusters to remove tSZ contamination, and iii) using a profile bias-hardened quadratic estimator for the lensing reconstruction.

We show first that on two sets of microwave sky simulations, \websky\ \citep{websky} and \sehgal\ \citep{Sehgal_2010}, the predicted level of bias to the estimated CMB lensing power spectrum is well below our statistical uncertainties. For the baseline analysis, with the MV estimator, the size of the fractional bias is below 1\% for most of the fiducial range of scales, $L$, used; the  bias to the inferred lensing amplitude, $\Alens$, is below $0.2\sigma$. When extending to higher $L$, foreground biases become more significant, but the bias to $\Alens$ remains at below $0.3\sigma$ for $L_{\mathrm{max}}=1300$.

In addition we present null tests performed on the DR6 data that leverage the frequency dependence of extragalactic foregrounds, and thus do not depend on having realistic microwave sky simulations. We investigate three ``lensing'' power spectra where the CMB lensing signal is nulled by differencing the f090 and f150 data both at the map level and the bandpower level, exploiting different sensitivities to the primary bispectrum, secondary bispectrum and trispectrum foreground components. All of these null tests pass (with $p$-value $\geq0.05$).

Finally, we demonstrate that using  CIB-deprojected maps in our lensing estimation produces lensing power spectrum bandpowers that are consistent with our baseline measurements, implying that CIB contamination is not likely to be a significant contaminant in the DR6 measurement. We note here a further test presented in our companion paper \citet{dr6-lensing-auto}, which is the consistency with the baseline measurement of the shear estimator of \citet{schaan19} and \citet{qu22}; this estimator uses only the quadrupolar contribution to the CMB mode-coupling induced by lensing. While the lensing power spectrum measurement with the  shear estimator is somewhat noiser than the baseline measurement, it is very encouraging that \citet{qu22} find the difference in the bandpowers is consistent with zero, with $\Delta\Alens=0.01\pm0.05$.

It is worth commenting here on the use of the DR6 lensing reconstruction maps for cross-correlation studies. The foreground bias estimates for the lensing auto-spectrum provided here are not directly applicable to cross-correlations of the lensing reconstructions with, e.g., maps of galaxy overdensity. These cross-correlations are impacted by biases analogous to the primary bispectrum bias described in \sect{sec:fg_biases}, due to the correlation between galaxy overdensity and CMB foregrounds that also trace the large-scale structure, especially the CIB and tSZ. The size of the contamination will depend on the specific tracer sample used for cross-correlation, but we do expect the mitigation strategies used here to also be very effective for these cross-correlations, as will be demonstrated for the case of unWISE galaxies in \citet{dr6-unwise}, and CMASS galaxies in \citet{dr6-cmass-eg}.

While polarisation data will become increasingly important for upcoming Simons Observatory (SO) lensing analyses, much of the $S/N$ will still depend on CMB temperature data, so careful treatment of extragalactic foregrounds will be required. With additional frequency channels at high resolution, as will be provided by SO, deprojecting both tSZ and CIB  could be more fruitful, including, for example, partial deprojection or composite approaches explored in \citet{abylkairov21}, \citet{sailer21} and \citet{darwish21a}. While deeper upcoming data from e.g. SO will demand greater control of foreground biases (given the reduced statistical uncertainties), it  will also allow fainter point-sources, dusty galaxies, and clusters to be detected and modeled or masked, although care must be taken not to introduce selection biases by preferentially masking higher convergence regions of the sky \citep{lembo22}.

\section*{Acknowledgements}

Support for ACT was through the U.S.~National Science Foundation through awards AST-0408698, AST-0965625, and AST-1440226 for the ACT project, as well as awards PHY-0355328, PHY-0855887 and PHY-1214379. Funding was also provided by Princeton University, the University of Pennsylvania, and a Canada Foundation for Innovation (CFI) award to UBC. ACT operated in the Parque Astron\'omico Atacama in northern Chile under the auspices of the Agencia Nacional de Investigaci\'on y Desarrollo (ANID). The development of multichroic detectors and lenses was supported by NASA grants NNX13AE56G and NNX14AB58G. Detector research at NIST was supported by the NIST Innovations in Measurement Science program. 

NM, BDS, FQ, BB, IAC, GSF, DH acknowledge support from the European Research Council (ERC) under the European Union’s Horizon 2020 research and innovation programme (Grant agreement No. 851274). BDS further acknowledges support from an STFC Ernest Rutherford Fellowship.

Computing for ACT was performed using the Princeton Research Computing resources at Princeton University, the National Energy Research Scientific Computing Center (NERSC), and the Niagara supercomputer at the SciNet HPC Consortium.

Some computations were performed on the Niagara supercomputer at the SciNet HPC Consortium and the Symmetry cluster at the Perimeter Institute. SciNet is funded by the CFI under the auspices of Compute Canada, the Government of Ontario, the Ontario Research Fund--Research Excellence, and the University of Toronto.  

 BDS, FJQ, BB, IAC, GSF, NM, DH acknowledge support from the European Research Council (ERC) under the European Union’s Horizon 2020 research and innovation programme (Grant agreement No. 851274). BDS further acknowledges support from an STFC Ernest Rutherford Fellowship. EC acknowledges support from the European Research Council (ERC) under the European Union’s Horizon 2020 research and innovation programme (Grant agreement No. 849169).
KM acknowledges support from the National Research Foundation of South Africa. OD acknowledges support from a SNSF Eccellenza Professorial Fellowship (No. 186879). CS acknowledges support from the Agencia Nacional de Investigaci\'on y Desarrollo (ANID) through FONDECYT grant no.\ 11191125 and BASAL project FB210003.
IAC acknowledges support from Fundaci\'on Mauricio y Carlota Botton. MH acknowledges support from the National Research Foundation of South Africa (grant no. 137975). SN was supported by a grant from the
Simons Foundation (CCA 918271, PBL). NS acknowledges support from NSF Grant number AST-1907657. JRB acknowledges support from NSERC and CIFAR and the Canadian Digital Alliance. 
GAM is part of Fermi Research Alliance, LLC under Contract No. DE-AC02-07CH11359 with the U.S. Department of Energy, Office of Science, Office of High Energy Physics.
OD acknowledges support from a SNSF Eccellenza Professorial Fellowship (No. 186879).
JCH acknowledges support from NSF grant AST-2108536, NASA grants 21-ATP21-0129 and 22-ADAP22-0145, DOE grant DE-SC00233966, the Sloan Foundation, and the Simons Foundation.
TN acknowledges support from JSPS KAKENHI (Grant No.\ JP20H05859 and No.\ JP22K03682) and World Premier International Research Center Initiative (WPI), MEXT, Japan.




\bibliography{refs}

\begin{thebibliography}{}
\expandafter\ifx\csname natexlab\endcsname\relax\def\natexlab#1{#1}\fi
\providecommand{\url}[1]{\href{#1}{#1}}
\providecommand{\dodoi}[1]{doi:~\href{http://doi.org/#1}{\nolinkurl{#1}}}
\providecommand{\doeprint}[1]{\href{http://ascl.net/#1}{\nolinkurl{http://ascl.net/#1}}}
\providecommand{\doarXiv}[1]{\href{https://arxiv.org/abs/#1}{\nolinkurl{https://arxiv.org/abs/#1}}}

\bibitem[{{Abazajian} {et~al.}(2016){Abazajian}, {Adshead}, {Ahmed}, {Allen},
  {Alonso}, {Arnold}, {Baccigalupi}, {Bartlett}, {Battaglia}, {Benson},
  {Bischoff}, {Borrill}, {Buza}, {Calabrese}, {Caldwell}, {Carlstrom}, {Chang},
  {Crawford}, {Cyr-Racine}, {De Bernardis}, {de Haan}, {di Serego Alighieri},
  {Dunkley}, {Dvorkin}, {Errard}, {Fabbian}, {Feeney}, {Ferraro}, {Filippini},
  {Flauger}, {Fuller}, {Gluscevic}, {Green}, {Grin}, {Grohs}, {Henning},
  {Hill}, {Hlozek}, {Holder}, {Holzapfel}, {Hu}, {Huffenberger}, {Keskitalo},
  {Knox}, {Kosowsky}, {Kovac}, {Kovetz}, {Kuo}, {Kusaka}, {Le Jeune}, {Lee},
  {Lilley}, {Loverde}, {Madhavacheril}, {Mantz}, {Marsh}, {McMahon},
  {Meerburg}, {Meyers}, {Miller}, {Munoz}, {Nguyen}, {Niemack}, {Peloso},
  {Peloton}, {Pogosian}, {Pryke}, {Raveri}, {Reichardt}, {Rocha}, {Rotti},
  {Schaan}, {Schmittfull}, {Scott}, {Sehgal}, {Shandera}, {Sherwin}, {Smith},
  {Sorbo}, {Starkman}, {Story}, {van Engelen}, {Vieira}, {Watson}, {Whitehorn},
  \& {Kimmy Wu}}]{Abazajian16}
{Abazajian}, K.~N., {Adshead}, P., {Ahmed}, Z., {et~al.} 2016, arXiv e-prints,
  arXiv:1610.02743, \dodoi{10.48550/arXiv.1610.02743}

\bibitem[{{Abylkairov} {et~al.}(2021){Abylkairov}, {Darwish}, {Hill}, \&
  {Sherwin}}]{abylkairov21}
{Abylkairov}, Y.~S., {Darwish}, O., {Hill}, J.~C., \& {Sherwin}, B.~D. 2021,
  \prd, 103, 103510, \dodoi{10.1103/PhysRevD.103.103510}

\bibitem[{{Arnaud} {et~al.}(2010){Arnaud}, {Pratt}, {Piffaretti},
  {B{\"o}hringer}, {Croston}, \& {Pointecouteau}}]{arnaud10}
{Arnaud}, M., {Pratt}, G.~W., {Piffaretti}, R., {et~al.} 2010, \aap, 517, A92,
  \dodoi{10.1051/0004-6361/200913416}

\bibitem[{{Atkins} {et~al.}(2023)}]{dr6-noise}
{Atkins}, Z., {et~al.} 2023, To be submitted to MNRAS

\bibitem[{{Battaglia} {et~al.}(2012){Battaglia}, {Bond}, {Pfrommer}, \&
  {Sievers}}]{battaglia12}
{Battaglia}, N., {Bond}, J.~R., {Pfrommer}, C., \& {Sievers}, J.~L. 2012, \apj,
  758, 75, \dodoi{10.1088/0004-637X/758/2/75}

\bibitem[{{Beck} {et~al.}(2020){Beck}, {Errard}, \& {Stompor}}]{beck20}
{Beck}, D., {Errard}, J., \& {Stompor}, R. 2020, \jcap, 2020, 030,
  \dodoi{10.1088/1475-7516/2020/06/030}

\bibitem[{{Bode} {et~al.}(2007){Bode}, {Ostriker}, {Weller}, \&
  {Shaw}}]{bode2007}
{Bode}, P., {Ostriker}, J.~P., {Weller}, J., \& {Shaw}, L. 2007, \apj, 663,
  139, \dodoi{10.1086/518432}

\bibitem[{{Bucher} \& {Louis}(2012)}]{bucher12}
{Bucher}, M., \& {Louis}, T. 2012, \mnras, 424, 1694,
  \dodoi{10.1111/j.1365-2966.2012.21138.x}

\bibitem[{{Cai} {et~al.}(2022){Cai}, {Madhavacheril}, {Hill}, \&
  {Kosowsky}}]{cai22}
{Cai}, H., {Madhavacheril}, M.~S., {Hill}, J.~C., \& {Kosowsky}, A. 2022, \prd,
  105, 043516, \dodoi{10.1103/PhysRevD.105.043516}

\bibitem[{{Carron} {et~al.}(2022){Carron}, {Mirmelstein}, \&
  {Lewis}}]{carron22}
{Carron}, J., {Mirmelstein}, M., \& {Lewis}, A. 2022, \jcap, 2022, 039,
  \dodoi{10.1088/1475-7516/2022/09/039}

\bibitem[{{Challinor} {et~al.}(2018){Challinor}, {Allison}, {Carron}, {Errard},
  {Feeney}, {Kitching}, {Lesgourgues}, {Lewis}, {Zubeld{\'\i}a}, {Achucarro},
  {Ade}, {Ashdown}, {Ballardini}, {Banday}, {Banerji}, {Bartlett}, {Bartolo},
  {Basak}, {Baumann}, {Bersanelli}, {Bonaldi}, {Bonato}, {Borrill}, {Bouchet},
  {Boulanger}, {Brinckmann}, {Bucher}, {Burigana}, {Buzzelli}, {Cai}, {Calvo},
  {Carvalho}, {Castellano}, {Chluba}, {Clesse}, {Colantoni}, {Coppolecchia},
  {Crook}, {d'Alessandro}, {de Bernardis}, {de Gasperis}, {De Zotti},
  {Delabrouille}, {Di Valentino}, {Diego}, {Fernandez-Cobos}, {Ferraro},
  {Finelli}, {Forastieri}, {Galli}, {Genova-Santos}, {Gerbino},
  {Gonz{\'a}lez-Nuevo}, {Grandis}, {Greenslade}, {Hagstotz}, {Hanany},
  {Handley}, {Hernandez-Monteagudo}, {Herv{\'\i}as-Caimapo}, {Hills}, {Hivon},
  {Kiiveri}, {Kisner}, {Kunz}, {Kurki-Suonio}, {Lamagna}, {Lasenby},
  {Lattanzi}, {Liguori}, {Lindholm}, {L{\'o}pez-Caniego}, {Luzzi}, {Maffei},
  {Martinez-Gonz{\'a}lez}, {Martins}, {Masi}, {Matarrese}, {McCarthy},
  {Melchiorri}, {Melin}, {Molinari}, {Monfardini}, {Natoli}, {Negrello},
  {Notari}, {Paiella}, {Paoletti}, {Patanchon}, {Piat}, {Pisano}, {Polastri},
  {Polenta}, {Pollo}, {Poulin}, {Quartin}, {Remazeilles}, {Roman},
  {Rubino-Martin}, {Salvati}, {Tartari}, {Tomasi}, {Tramonte}, {Trappe},
  {Trombetti}, {Tucker}, {Valiviita}, {Van de Weijgaert}, {van Tent}, {Vennin},
  {Vielva}, {Vittorio}, {Young}, \& {Zannoni}}]{challinor18}
{Challinor}, A., {Allison}, R., {Carron}, J., {et~al.} 2018, \jcap, 2018, 018,
  \dodoi{10.1088/1475-7516/2018/04/018}

\bibitem[{{Coulton} {et~al.}(in prep.)}]{dr6-ilc}
{Coulton}, W., {et~al.} in prep., To be submitted to MNRAS

\bibitem[{{Darwish} {et~al.}(2021{\natexlab{a}}){Darwish}, {Sherwin}, {Sailer},
  {Schaan}, \& {Ferraro}}]{darwish21a}
{Darwish}, O., {Sherwin}, B.~D., {Sailer}, N., {Schaan}, E., \& {Ferraro}, S.
  2021{\natexlab{a}}, arXiv e-prints, arXiv:2111.00462.
\newblock \doarXiv{2111.00462}

\bibitem[{{Darwish} {et~al.}(2021{\natexlab{b}}){Darwish}, {Madhavacheril},
  {Sherwin}, {Aiola}, {Battaglia}, {Beall}, {Becker}, {Bond}, {Calabrese},
  {Choi}, {Devlin}, {Dunkley}, {D{\"u}nner}, {Ferraro}, {Fox}, {Gallardo},
  {Guan}, {Halpern}, {Han}, {Hasselfield}, {Hill}, {Hilton}, {Hilton},
  {Hincks}, {Patty Ho}, {Hubmayr}, {Hughes}, {Koopman}, {Kosowsky}, {Van
  Lanen}, {Louis}, {Lungu}, {MacInnis}, {Maurin}, {McMahon}, {Moodley},
  {Naess}, {Namikawa}, {Nati}, {Newburgh}, {Nibarger}, {Niemack}, {Page},
  {Partridge}, {Qu}, {Robertson}, {Schillaci}, {Schmitt}, {Sehgal},
  {Sif{\'o}n}, {Spergel}, {Staggs}, {Storer}, {van Engelen}, \&
  {Wollack}}]{darwish21}
{Darwish}, O., {Madhavacheril}, M.~S., {Sherwin}, B.~D., {et~al.}
  2021{\natexlab{b}}, \mnras, 500, 2250, \dodoi{10.1093/mnras/staa3438}

\bibitem[{{Das} {et~al.}(2011){Das}, {Sherwin}, {Aguirre}, {Appel}, {Bond},
  {Carvalho}, {Devlin}, {Dunkley}, {D{\"u}nner}, {Essinger-Hileman}, {Fowler},
  {Hajian}, {Halpern}, {Hasselfield}, {Hincks}, {Hlozek}, {Huffenberger},
  {Hughes}, {Irwin}, {Klein}, {Kosowsky}, {Lupton}, {Marriage}, {Marsden},
  {Menanteau}, {Moodley}, {Niemack}, {Nolta}, {Page}, {Parker}, {Reese},
  {Schmitt}, {Sehgal}, {Sievers}, {Spergel}, {Staggs}, {Swetz}, {Switzer},
  {Thornton}, {Visnjic}, \& {Wollack}}]{das11}
{Das}, S., {Sherwin}, B.~D., {Aguirre}, P., {et~al.} 2011, \prl, 107, 021301,
  \dodoi{10.1103/PhysRevLett.107.021301}

\bibitem[{{Delabrouille} {et~al.}(2009){Delabrouille}, {Cardoso}, {Le Jeune},
  {Betoule}, {Fay}, \& {Guilloux}}]{delabrouille09}
{Delabrouille}, J., {Cardoso}, J.~F., {Le Jeune}, M., {et~al.} 2009, \aap, 493,
  835, \dodoi{10.1051/0004-6361:200810514}

\bibitem[{{Dunkley} {et~al.}(2013){Dunkley}, {Calabrese}, {Sievers}, {Addison},
  {Battaglia}, {Battistelli}, {Bond}, {Das}, {Devlin}, {D{\"u}nner}, {Fowler},
  {Gralla}, {Hajian}, {Halpern}, {Hasselfield}, {Hincks}, {Hlozek}, {Hughes},
  {Irwin}, {Kosowsky}, {Louis}, {Marriage}, {Marsden}, {Menanteau}, {Moodley},
  {Niemack}, {Nolta}, {Page}, {Partridge}, {Sehgal}, {Spergel}, {Staggs},
  {Switzer}, {Trac}, \& {Wollack}}]{dunkley13}
{Dunkley}, J., {Calabrese}, E., {Sievers}, J., {et~al.} 2013, \jcap, 2013, 025,
  \dodoi{10.1088/1475-7516/2013/07/025}

\bibitem[{{Farren} {et~al.}(in prep.)}]{dr6-unwise}
{Farren}, G., {et~al.} in prep., To be submitted to MNRAS

\bibitem[{{Ferraro} \& {Hill}(2018)}]{ferraro18}
{Ferraro}, S., \& {Hill}, J.~C. 2018, \prd, 97, 023512,
  \dodoi{10.1103/PhysRevD.97.023512}

\bibitem[{{Haehnelt} \& {Tegmark}(1996)}]{haehnelt2006}
{Haehnelt}, M.~G., \& {Tegmark}, M. 1996, \mnras, 279, 545,
  \dodoi{10.1093/mnras/279.2.545}

\bibitem[{{Hanson} {et~al.}(2011){Hanson}, {Challinor}, {Efstathiou}, \&
  {Bielewicz}}]{hanson11}
{Hanson}, D., {Challinor}, A., {Efstathiou}, G., \& {Bielewicz}, P. 2011, \prd,
  83, 043005, \dodoi{10.1103/PhysRevD.83.043005}

\bibitem[{Hilton {et~al.}(2021)Hilton, Sifón, Naess, Madhavacheril, Oguri,
  Rozo, Rykoff, Abbott, Adhikari, Aguena, \& et~al.}]{Hilton_2021}
Hilton, M., Sifón, C., Naess, S., {et~al.} 2021, The Astrophysical Journal
  Supplement Series, 253, 3, \dodoi{10.3847/1538-4365/abd023}

\bibitem[{{Hirata} {et~al.}(2008){Hirata}, {Ho}, {Padmanabhan}, {Seljak}, \&
  {Bahcall}}]{hirata08}
{Hirata}, C.~M., {Ho}, S., {Padmanabhan}, N., {Seljak}, U., \& {Bahcall}, N.~A.
  2008, \prd, 78, 043520, \dodoi{10.1103/PhysRevD.78.043520}

\bibitem[{{Hu} {et~al.}(2007){Hu}, {DeDeo}, \& {Vale}}]{hu07}
{Hu}, W., {DeDeo}, S., \& {Vale}, C. 2007, New Journal of Physics, 9, 441,
  \dodoi{10.1088/1367-2630/9/12/441}

\bibitem[{{Hu} \& {Okamoto}(2002)}]{huokamoto2002}
{Hu}, W., \& {Okamoto}, T. 2002, \apj, 574, 566, \dodoi{10.1086/341110}

\bibitem[{{Lembo} {et~al.}(2022){Lembo}, {Fabbian}, {Carron}, \&
  {Lewis}}]{lembo22}
{Lembo}, M., {Fabbian}, G., {Carron}, J., \& {Lewis}, A. 2022, \prd, 106,
  023525, \dodoi{10.1103/PhysRevD.106.023525}

\bibitem[{{Lewis} \& {Challinor}(2006)}]{lewis2006}
{Lewis}, A., \& {Challinor}, A. 2006, \physrep, 429, 1,
  \dodoi{10.1016/j.physrep.2006.03.002}

\bibitem[{{Li} {et~al.}(2022){Li}, {Puglisi}, {Madhavacheril}, \&
  {Alvarez}}]{Li22}
{Li}, Z., {Puglisi}, G., {Madhavacheril}, M.~S., \& {Alvarez}, M.~A. 2022,
  \jcap, 2022, 029, \dodoi{10.1088/1475-7516/2022/08/029}

\bibitem[{{Madhavacheril} {et~al.}(2023)}]{dr6-lensing-cosmo}
{Madhavacheril}, M., {et~al.} 2023, To be submitted to ApJ

\bibitem[{{Madhavacheril} \& {Hill}(2018)}]{madhavacheril18}
{Madhavacheril}, M.~S., \& {Hill}, J.~C. 2018, \prd, 98, 023534,
  \dodoi{10.1103/PhysRevD.98.023534}

\bibitem[{{Madhavacheril} {et~al.}(2020{\natexlab{a}}){Madhavacheril}, {Smith},
  {Sherwin}, \& {Naess}}]{madhavacheril20b}
{Madhavacheril}, M.~S., {Smith}, K.~M., {Sherwin}, B.~D., \& {Naess}, S.
  2020{\natexlab{a}}, arXiv e-prints, arXiv:2011.02475,
  \dodoi{10.48550/arXiv.2011.02475}

\bibitem[{{Madhavacheril} {et~al.}(2020{\natexlab{b}}){Madhavacheril}, {Hill},
  {N{\ae}ss}, {Addison}, {Aiola}, {Baildon}, {Battaglia}, {Bean}, {Bond},
  {Calabrese}, {Calafut}, {Choi}, {Darwish}, {Datta}, {Devlin}, {Dunkley},
  {D{\"u}nner}, {Ferraro}, {Gallardo}, {Gluscevic}, {Halpern}, {Han},
  {Hasselfield}, {Hilton}, {Hincks}, {Hlo{\v{z}}ek}, {Ho}, {Huffenberger},
  {Hughes}, {Koopman}, {Kosowsky}, {Lokken}, {Louis}, {Lungu}, {MacInnis},
  {Maurin}, {McMahon}, {Moodley}, {Nati}, {Niemack}, {Page}, {Partridge},
  {Robertson}, {Sehgal}, {Schaan}, {Schillaci}, {Sherwin}, {Sif{\'o}n},
  {Simon}, {Spergel}, {Staggs}, {Storer}, {van Engelen}, {Vavagiakis},
  {Wollack}, \& {Xu}}]{madhavacheril20}
{Madhavacheril}, M.~S., {Hill}, J.~C., {N{\ae}ss}, S., {et~al.}
  2020{\natexlab{b}}, \prd, 102, 023534, \dodoi{10.1103/PhysRevD.102.023534}

\bibitem[{{Naess} {et~al.}(2020){Naess}, {Aiola}, {Austermann}, {Battaglia},
  {Beall}, {Becker}, {Bond}, {Calabrese}, {Choi}, {Cothard}, {Crowley},
  {Darwish}, {Datta}, {Denison}, {Devlin}, {Duell}, {Duff}, {Duivenvoorden},
  {Dunkley}, {D{\"u}nner}, {Fox}, {Gallardo}, {Halpern}, {Han}, {Hasselfield},
  {Hill}, {Hilton}, {Hilton}, {Hincks}, {Hlo{\v{z}}ek}, {Ho}, {Hubmayr},
  {Huffenberger}, {Hughes}, {Kosowsky}, {Louis}, {Madhavacheril}, {McMahon},
  {Moodley}, {Nati}, {Nibarger}, {Niemack}, {Page}, {Partridge}, {Salatino},
  {Schaan}, {Schillaci}, {Schmitt}, {Sherwin}, {Sehgal}, {Sif{\'o}n},
  {Spergel}, {Staggs}, {Stevens}, {Storer}, {Ullom}, {Vale}, {Van Engelen},
  {Van Lanen}, {Vavagiakis}, {Wollack}, \& {Xu}}]{naess20}
{Naess}, S., {Aiola}, S., {Austermann}, J.~E., {et~al.} 2020, \jcap, 2020, 046,
  \dodoi{10.1088/1475-7516/2020/12/046}

\bibitem[{{Naess} {et~al.}(in prep.)}]{dr6-maps}
{Naess}, S., {et~al.} in prep., To be submitted to MNRAS

\bibitem[{{Namikawa} {et~al.}(2013){Namikawa}, {Hanson}, \&
  {Takahashi}}]{namikawa13}
{Namikawa}, T., {Hanson}, D., \& {Takahashi}, R. 2013, \mnras, 431, 609,
  \dodoi{10.1093/mnras/stt195}

\bibitem[{{Namikawa} \& {Takahashi}(2014)}]{namikawa14}
{Namikawa}, T., \& {Takahashi}, R. 2014, \mnras, 438, 1507,
  \dodoi{10.1093/mnras/stt2290}

\bibitem[{{Osborne} {et~al.}(2014){Osborne}, {Hanson}, \&
  {Dor{\'e}}}]{osborne14}
{Osborne}, S.~J., {Hanson}, D., \& {Dor{\'e}}, O. 2014, \jcap, 2014, 024,
  \dodoi{10.1088/1475-7516/2014/03/024}

\bibitem[{{Planck Collaboration} {et~al.}(2016){Planck Collaboration}, {Ade},
  {Aghanim}, {Arg{\"u}eso}, {Arnaud}, {Ashdown}, {Aumont}, {Baccigalupi},
  {Banday}, {Barreiro}, {Bartolo}, {Battaner}, {Beichman}, {Benabed},
  {Beno{\^\i}t}, {Benoit-L{\'e}vy}, {Bernard}, {Bersanelli}, {Bielewicz},
  {Bock}, {B{\"o}hringer}, {Bonaldi}, {Bonavera}, {Bond}, {Borrill}, {Bouchet},
  {Boulanger}, {Bucher}, {Burigana}, {Butler}, {Calabrese}, {Cardoso},
  {Carvalho}, {Catalano}, {Challinor}, {Chamballu}, {Chary}, {Chiang},
  {Christensen}, {Clemens}, {Clements}, {Colombi}, {Colombo}, {Combet},
  {Couchot}, {Coulais}, {Crill}, {Curto}, {Cuttaia}, {Danese}, {Davies},
  {Davis}, {de Bernardis}, {de Rosa}, {de Zotti}, {Delabrouille}, {D{\'e}sert},
  {Dickinson}, {Diego}, {Dole}, {Donzelli}, {Dor{\'e}}, {Douspis}, {Ducout},
  {Dupac}, {Efstathiou}, {Elsner}, {En{\ss}lin}, {Eriksen}, {Falgarone},
  {Fergusson}, {Finelli}, {Forni}, {Frailis}, {Fraisse}, {Franceschi},
  {Frejsel}, {Galeotta}, {Galli}, {Ganga}, {Giard}, {Giraud-H{\'e}raud},
  {Gjerl{\o}w}, {Gonz{\'a}lez-Nuevo}, {G{\'o}rski}, {Gratton}, {Gregorio},
  {Gruppuso}, {Gudmundsson}, {Hansen}, {Hanson}, {Harrison}, {Helou},
  {Henrot-Versill{\'e}}, {Hern{\'a}ndez-Monteagudo}, {Herranz}, {Hildebrandt},
  {Hivon}, {Hobson}, {Holmes}, {Hornstrup}, {Hovest}, {Huffenberger}, {Hurier},
  {Jaffe}, {Jaffe}, {Jones}, {Juvela}, {Keih{\"a}nen}, {Keskitalo}, {Kisner},
  {Kneissl}, {Knoche}, {Kunz}, {Kurki-Suonio}, {Lagache},
  {L{\"a}hteenm{\"a}ki}, {Lamarre}, {Lasenby}, {Lattanzi}, {Lawrence}, {Leahy},
  {Leonardi}, {Le{\'o}n-Tavares}, {Lesgourgues}, {Levrier}, {Liguori}, {Lilje},
  {Linden-V{\o}rnle}, {L{\'o}pez-Caniego}, {Lubin}, {Mac{\'\i}as-P{\'e}rez},
  {Maggio}, {Maino}, {Mandolesi}, {Mangilli}, {Maris}, {Marshall}, {Martin},
  {Mart{\'\i}nez-Gonz{\'a}lez}, {Masi}, {Matarrese}, {McGehee}, {Meinhold},
  {Melchiorri}, {Mendes}, {Mennella}, {Migliaccio}, {Mitra},
  {Miville-Desch{\^e}nes}, {Moneti}, {Montier}, {Morgante}, {Mortlock}, {Moss},
  {Munshi}, {Murphy}, {Naselsky}, {Nati}, {Natoli}, {Negrello}, {Netterfield},
  {N{\o}rgaard-Nielsen}, {Noviello}, {Novikov}, {Novikov}, {Oxborrow}, {Paci},
  {Pagano}, {Pajot}, {Paladini}, {Paoletti}, {Partridge}, {Pasian},
  {Patanchon}, {Pearson}, {Perdereau}, {Perotto}, {Perrotta}, {Pettorino},
  {Piacentini}, {Piat}, {Pierpaoli}, {Pietrobon}, {Plaszczynski},
  {Pointecouteau}, {Polenta}, {Pratt}, {Pr{\'e}zeau}, {Prunet}, {Puget},
  {Rachen}, {Reach}, {Rebolo}, {Reinecke}, {Remazeilles}, {Renault}, {Renzi},
  {Ristorcelli}, {Rocha}, {Rosset}, {Rossetti}, {Roudier}, {Rowan-Robinson},
  {Rubi{\~n}o-Mart{\'\i}n}, {Rusholme}, {Sandri}, {Sanghera}, {Santos},
  {Savelainen}, {Savini}, {Scott}, {Seiffert}, {Shellard}, {Spencer},
  {Stolyarov}, {Sudiwala}, {Sunyaev}, {Sutton}, {Suur-Uski}, {Sygnet},
  {Tauber}, {Terenzi}, {Toffolatti}, {Tomasi}, {Tornikoski}, {Tristram},
  {Tucci}, {Tuovinen}, {T{\"u}rler}, {Umana}, {Valenziano}, {Valiviita}, {Van
  Tent}, {Vielva}, {Villa}, {Wade}, {Walter}, {Wandelt}, {Wehus}, {Yvon},
  {Zacchei}, \& {Zonca}}]{ade16}
{Planck Collaboration}, {Ade}, P.~A.~R., {Aghanim}, N., {et~al.} 2016, \aap,
  594, A26, \dodoi{10.1051/0004-6361/201526914}

\bibitem[{{Planck Collaboration} {et~al.}(2020{\natexlab{a}}){Planck
  Collaboration}, {Aghanim}, {Akrami}, {Ashdown}, {Aumont}, {Baccigalupi},
  {Ballardini}, {Banday}, {Barreiro}, {Bartolo}, {Basak}, {Benabed}, {Bernard},
  {Bersanelli}, {Bielewicz}, {Bock}, {Bond}, {Borrill}, {Bouchet}, {Boulanger},
  {Bucher}, {Burigana}, {Calabrese}, {Cardoso}, {Carron}, {Challinor},
  {Chiang}, {Colombo}, {Combet}, {Crill}, {Cuttaia}, {de Bernardis}, {de
  Zotti}, {Delabrouille}, {Di Valentino}, {Diego}, {Dor{\'e}}, {Douspis},
  {Ducout}, {Dupac}, {Efstathiou}, {Elsner}, {En{\ss}lin}, {Eriksen},
  {Fantaye}, {Fernandez-Cobos}, {Finelli}, {Forastieri}, {Frailis}, {Fraisse},
  {Franceschi}, {Frolov}, {Galeotta}, {Galli}, {Ganga}, {G{\'e}nova-Santos},
  {Gerbino}, {Ghosh}, {Gonz{\'a}lez-Nuevo}, {G{\'o}rski}, {Gratton},
  {Gruppuso}, {Gudmundsson}, {Hamann}, {Handley}, {Hansen}, {Herranz}, {Hivon},
  {Huang}, {Jaffe}, {Jones}, {Karakci}, {Keih{\"a}nen}, {Keskitalo}, {Kiiveri},
  {Kim}, {Knox}, {Krachmalnicoff}, {Kunz}, {Kurki-Suonio}, {Lagache},
  {Lamarre}, {Lasenby}, {Lattanzi}, {Lawrence}, {Le Jeune}, {Levrier}, {Lewis},
  {Liguori}, {Lilje}, {Lindholm}, {L{\'o}pez-Caniego}, {Lubin}, {Ma},
  {Mac{\'\i}as-P{\'e}rez}, {Maggio}, {Maino}, {Mandolesi}, {Mangilli},
  {Marcos-Caballero}, {Maris}, {Martin}, {Mart{\'\i}nez-Gonz{\'a}lez},
  {Matarrese}, {Mauri}, {McEwen}, {Melchiorri}, {Mennella}, {Migliaccio},
  {Miville-Desch{\^e}nes}, {Molinari}, {Moneti}, {Montier}, {Morgante}, {Moss},
  {Natoli}, {Pagano}, {Paoletti}, {Partridge}, {Patanchon}, {Perrotta},
  {Pettorino}, {Piacentini}, {Polastri}, {Polenta}, {Puget}, {Rachen},
  {Reinecke}, {Remazeilles}, {Renzi}, {Rocha}, {Rosset}, {Roudier},
  {Rubi{\~n}o-Mart{\'\i}n}, {Ruiz-Granados}, {Salvati}, {Sandri}, {Savelainen},
  {Scott}, {Sirignano}, {Sunyaev}, {Suur-Uski}, {Tauber}, {Tavagnacco},
  {Tenti}, {Toffolatti}, {Tomasi}, {Trombetti}, {Valiviita}, {Van Tent},
  {Vielva}, {Villa}, {Vittorio}, {Wandelt}, {Wehus}, {White}, {White},
  {Zacchei}, \& {Zonca}}]{plancklensing2018}
{Planck Collaboration}, {Aghanim}, N., {Akrami}, Y., {et~al.}
  2020{\natexlab{a}}, \aap, 641, A8, \dodoi{10.1051/0004-6361/201833886}

\bibitem[{{Planck Collaboration} {et~al.}(2020{\natexlab{b}}){Planck
  Collaboration}, {Akrami}, {Ashdown}, {Aumont}, {Baccigalupi}, {Ballardini},
  {Banday}, {Barreiro}, {Bartolo}, {Basak}, {Benabed}, {Bersanelli},
  {Bielewicz}, {Bond}, {Borrill}, {Bouchet}, {Boulanger}, {Bucher}, {Burigana},
  {Calabrese}, {Cardoso}, {Carron}, {Casaponsa}, {Challinor}, {Colombo},
  {Combet}, {Crill}, {Cuttaia}, {de Bernardis}, {de Rosa}, {de Zotti},
  {Delabrouille}, {Delouis}, {Di Valentino}, {Dickinson}, {Diego}, {Donzelli},
  {Dor{\'e}}, {Ducout}, {Dupac}, {Efstathiou}, {Elsner}, {En{\ss}lin},
  {Eriksen}, {Falgarone}, {Fernandez-Cobos}, {Finelli}, {Forastieri},
  {Frailis}, {Fraisse}, {Franceschi}, {Frolov}, {Galeotta}, {Galli}, {Ganga},
  {G{\'e}nova-Santos}, {Gerbino}, {Ghosh}, {Gonz{\'a}lez-Nuevo}, {G{\'o}rski},
  {Gratton}, {Gruppuso}, {Gudmundsson}, {Handley}, {Hansen}, {Helou},
  {Herranz}, {Hildebrandt}, {Huang}, {Jaffe}, {Karakci}, {Keih{\"a}nen},
  {Keskitalo}, {Kiiveri}, {Kim}, {Kisner}, {Krachmalnicoff}, {Kunz},
  {Kurki-Suonio}, {Lagache}, {Lamarre}, {Lasenby}, {Lattanzi}, {Lawrence}, {Le
  Jeune}, {Levrier}, {Liguori}, {Lilje}, {Lindholm}, {L{\'o}pez-Caniego},
  {Lubin}, {Ma}, {Mac{\'\i}as-P{\'e}rez}, {Maggio}, {Maino}, {Mandolesi},
  {Mangilli}, {Marcos-Caballero}, {Maris}, {Martin},
  {Mart{\'\i}nez-Gonz{\'a}lez}, {Matarrese}, {Mauri}, {McEwen}, {Meinhold},
  {Melchiorri}, {Mennella}, {Migliaccio}, {Miville-Desch{\^e}nes}, {Molinari},
  {Moneti}, {Montier}, {Morgante}, {Natoli}, {Oppizzi}, {Pagano}, {Paoletti},
  {Partridge}, {Peel}, {Pettorino}, {Piacentini}, {Polenta}, {Puget}, {Rachen},
  {Reinecke}, {Remazeilles}, {Renzi}, {Rocha}, {Roudier},
  {Rubi{\~n}o-Mart{\'\i}n}, {Ruiz-Granados}, {Salvati}, {Sandri}, {Savelainen},
  {Scott}, {Seljebotn}, {Sirignano}, {Spencer}, {Suur-Uski}, {Tauber},
  {Tavagnacco}, {Tenti}, {Thommesen}, {Toffolatti}, {Tomasi}, {Trombetti},
  {Valiviita}, {Van Tent}, {Vielva}, {Villa}, {Vittorio}, {Wandelt}, {Wehus},
  {Zacchei}, \& {Zonca}}]{planck2018maps}
{Planck Collaboration}, {Akrami}, Y., {Ashdown}, M., {et~al.}
  2020{\natexlab{b}}, \aap, 641, A4, \dodoi{10.1051/0004-6361/201833881}

\bibitem[{{Planck Collaboration} {et~al.}(2020{\natexlab{c}}){Planck
  Collaboration}, {Akrami}, {Andersen}, {Ashdown}, {Baccigalupi}, {Ballardini},
  {Banday}, {Barreiro}, {Bartolo}, {Basak}, {Benabed}, {Bernard}, {Bersanelli},
  {Bielewicz}, {Bond}, {Borrill}, {Burigana}, {Butler}, {Calabrese},
  {Casaponsa}, {Chiang}, {Colombo}, {Combet}, {Crill}, {Cuttaia}, {de
  Bernardis}, {de Rosa}, {de Zotti}, {Delabrouille}, {Di Valentino}, {Diego},
  {Dor{\'e}}, {Douspis}, {Dupac}, {Eriksen}, {Fernandez-Cobos}, {Finelli},
  {Frailis}, {Fraisse}, {Franceschi}, {Frolov}, {Galeotta}, {Galli}, {Ganga},
  {Gerbino}, {Ghosh}, {Gonz{\'a}lez-Nuevo}, {G{\'o}rski}, {Gruppuso},
  {Gudmundsson}, {Handley}, {Helou}, {Herranz}, {Hildebrandt}, {Hivon},
  {Huang}, {Jaffe}, {Jones}, {Keih{\"a}nen}, {Keskitalo}, {Kiiveri}, {Kim},
  {Kisner}, {Krachmalnicoff}, {Kunz}, {Kurki-Suonio}, {Lasenby}, {Lattanzi},
  {Lawrence}, {Le Jeune}, {Levrier}, {Liguori}, {Lilje}, {Lilley}, {Lindholm},
  {L{\'o}pez-Caniego}, {Lubin}, {Mac{\'\i}as-P{\'e}rez}, {Maino}, {Mandolesi},
  {Marcos-Caballero}, {Maris}, {Martin}, {Mart{\'\i}nez-Gonz{\'a}lez},
  {Matarrese}, {Mauri}, {McEwen}, {Meinhold}, {Mennella}, {Migliaccio},
  {Mitra}, {Molinari}, {Montier}, {Morgante}, {Moss}, {Natoli}, {Paoletti},
  {Partridge}, {Patanchon}, {Pearson}, {Pearson}, {Perrotta}, {Piacentini},
  {Polenta}, {Rachen}, {Reinecke}, {Remazeilles}, {Renzi}, {Rocha}, {Rosset},
  {Roudier}, {Rubi{\~n}o-Mart{\'\i}n}, {Ruiz-Granados}, {Salvati},
  {Savelainen}, {Scott}, {Sirignano}, {Sirri}, {Spencer}, {Suur-Uski},
  {Svalheim}, {Tauber}, {Tavagnacco}, {Tenti}, {Terenzi}, {Thommesen},
  {Toffolatti}, {Tomasi}, {Tristram}, {Trombetti}, {Valiviita}, {Van Tent},
  {Vielva}, {Villa}, {Vittorio}, {Wandelt}, {Wehus}, {Zacchei}, \&
  {Zonca}}]{akrami20}
{Planck Collaboration}, {Akrami}, Y., {Andersen}, K.~J., {et~al.}
  2020{\natexlab{c}}, \aap, 643, A42, \dodoi{10.1051/0004-6361/202038073}

\bibitem[{{Planck Collaboration} {et~al.}(2020{\natexlab{d}}){Planck
  Collaboration}, {Akrami}, {Andersen}, {Ashdown}, {Baccigalupi}, {Ballardini},
  {Banday}, {Barreiro}, {Bartolo}, {Basak}, {Benabed}, {Bernard}, {Bersanelli},
  {Bielewicz}, {Bond}, {Borrill}, {Burigana}, {Butler}, {Calabrese},
  {Casaponsa}, {Chiang}, {Colombo}, {Combet}, {Crill}, {Cuttaia}, {de
  Bernardis}, {de Rosa}, {de Zotti}, {Delabrouille}, {Di Valentino}, {Diego},
  {Dor{\'e}}, {Douspis}, {Dupac}, {Eriksen}, {Fernandez-Cobos}, {Finelli},
  {Frailis}, {Fraisse}, {Franceschi}, {Frolov}, {Galeotta}, {Galli}, {Ganga},
  {Gerbino}, {Ghosh}, {Gonz{\'a}lez-Nuevo}, {G{\'o}rski}, {Gruppuso},
  {Gudmundsson}, {Handley}, {Helou}, {Herranz}, {Hildebrandt}, {Hivon},
  {Huang}, {Jaffe}, {Jones}, {Keih{\"a}nen}, {Keskitalo}, {Kiiveri}, {Kim},
  {Kisner}, {Krachmalnicoff}, {Kunz}, {Kurki-Suonio}, {Lasenby}, {Lattanzi},
  {Lawrence}, {Le Jeune}, {Levrier}, {Liguori}, {Lilje}, {Lilley}, {Lindholm},
  {L{\'o}pez-Caniego}, {Lubin}, {Mac{\'\i}as-P{\'e}rez}, {Maino}, {Mandolesi},
  {Marcos-Caballero}, {Maris}, {Martin}, {Mart{\'\i}nez-Gonz{\'a}lez},
  {Matarrese}, {Mauri}, {McEwen}, {Meinhold}, {Mennella}, {Migliaccio},
  {Mitra}, {Molinari}, {Montier}, {Morgante}, {Moss}, {Natoli}, {Paoletti},
  {Partridge}, {Patanchon}, {Pearson}, {Pearson}, {Perrotta}, {Piacentini},
  {Polenta}, {Rachen}, {Reinecke}, {Remazeilles}, {Renzi}, {Rocha}, {Rosset},
  {Roudier}, {Rubi{\~n}o-Mart{\'\i}n}, {Ruiz-Granados}, {Salvati},
  {Savelainen}, {Scott}, {Sirignano}, {Sirri}, {Spencer}, {Suur-Uski},
  {Svalheim}, {Tauber}, {Tavagnacco}, {Tenti}, {Terenzi}, {Thommesen},
  {Toffolatti}, {Tomasi}, {Tristram}, {Trombetti}, {Valiviita}, {Van Tent},
  {Vielva}, {Villa}, {Vittorio}, {Wandelt}, {Wehus}, {Zacchei}, \&
  {Zonca}}]{plancknpipe}
---. 2020{\natexlab{d}}, \aap, 643, A42, \dodoi{10.1051/0004-6361/202038073}

\bibitem[{{Qu} {et~al.}(2023)}]{dr6-lensing-auto}
{Qu}, F., {et~al.} 2023, To be submitted to ApJ

\bibitem[{{Qu} {et~al.}(2022){Qu}, {Challinor}, \& {Sherwin}}]{qu22}
{Qu}, F.~J., {Challinor}, A., \& {Sherwin}, B.~D. 2022, arXiv e-prints,
  arXiv:2208.14988.
\newblock \doarXiv{2208.14988}

\bibitem[{{Remazeilles} {et~al.}(2011){Remazeilles}, {Delabrouille}, \&
  {Cardoso}}]{remazeilles11}
{Remazeilles}, M., {Delabrouille}, J., \& {Cardoso}, J.-F. 2011, \mnras, 410,
  2481, \dodoi{10.1111/j.1365-2966.2010.17624.x}

\bibitem[{{Sailer} {et~al.}(2023){Sailer}, {Ferraro}, \& {Schaan}}]{sailer23b}
{Sailer}, N., {Ferraro}, S., \& {Schaan}, E. 2023, \prd, 107, 023504,
  \dodoi{10.1103/PhysRevD.107.02350410.48550/arXiv.2211.03786}

\bibitem[{{Sailer} {et~al.}(2020){Sailer}, {Schaan}, \& {Ferraro}}]{sailer20}
{Sailer}, N., {Schaan}, E., \& {Ferraro}, S. 2020, \prd, 102, 063517,
  \dodoi{10.1103/PhysRevD.102.063517}

\bibitem[{{Sailer} {et~al.}(2021){Sailer}, {Schaan}, {Ferraro}, {Darwish}, \&
  {Sherwin}}]{sailer21}
{Sailer}, N., {Schaan}, E., {Ferraro}, S., {Darwish}, O., \& {Sherwin}, B.
  2021, \prd, 104, 123514, \dodoi{10.1103/PhysRevD.104.123514}

\bibitem[{{Schaan} \& {Ferraro}(2019)}]{schaan19}
{Schaan}, E., \& {Ferraro}, S. 2019, \prl, 122, 181301,
  \dodoi{10.1103/PhysRevLett.122.181301}

\bibitem[{Sehgal {et~al.}(2010)Sehgal, Bode, Das, Hernandez-Monteagudo,
  Huffenberger, Lin, Ostriker, \& Trac}]{Sehgal_2010}
Sehgal, N., Bode, P., Das, S., {et~al.} 2010, The Astrophysical Journal, 709,
  920–936, \dodoi{10.1088/0004-637x/709/2/920}

\bibitem[{{Shang} {et~al.}(2012){Shang}, {Haiman}, {Knox}, \& {Oh}}]{shang12}
{Shang}, C., {Haiman}, Z., {Knox}, L., \& {Oh}, S.~P. 2012, \mnras, 421, 2832,
  \dodoi{10.1111/j.1365-2966.2012.20510.x}

\bibitem[{Smith {et~al.}(2007)Smith, Zahn, \& Dor\'e}]{smith07}
Smith, K.~M., Zahn, O., \& Dor\'e, O. 2007, Phys. Rev. D, 76, 043510,
  \dodoi{10.1103/PhysRevD.76.043510}

\bibitem[{{Staniszewski} {et~al.}(2009){Staniszewski}, {Ade}, {Aird}, {Benson},
  {Bleem}, {Carlstrom}, {Chang}, {Cho}, {Crawford}, {Crites}, {de Haan},
  {Dobbs}, {Halverson}, {Holder}, {Holzapfel}, {Hrubes}, {Joy}, {Keisler},
  {Lanting}, {Lee}, {Leitch}, {Loehr}, {Lueker}, {McMahon}, {Mehl}, {Meyer},
  {Mohr}, {Montroy}, {Ngeow}, {Padin}, {Plagge}, {Pryke}, {Reichardt}, {Ruhl},
  {Schaffer}, {Shaw}, {Shirokoff}, {Spieler}, {Stalder}, {Stark},
  {Vanderlinde}, {Vieira}, {Zahn}, \& {Zenteno}}]{Staniszewski2009}
{Staniszewski}, Z., {Ade}, P.~A.~R., {Aird}, K.~A., {et~al.} 2009, \apj, 701,
  32, \dodoi{10.1088/0004-637X/701/1/32}

\bibitem[{{Stein} {et~al.}(2019){Stein}, {Alvarez}, \& {Bond}}]{stein19}
{Stein}, G., {Alvarez}, M.~A., \& {Bond}, J.~R. 2019, \mnras, 483, 2236,
  \dodoi{10.1093/mnras/sty3226}

\bibitem[{{Stein} {et~al.}(2020){Stein}, {Alvarez}, {Bond}, {van Engelen}, \&
  {Battaglia}}]{websky}
{Stein}, G., {Alvarez}, M.~A., {Bond}, J.~R., {van Engelen}, A., \&
  {Battaglia}, N. 2020, \jcap, 2020, 012, \dodoi{10.1088/1475-7516/2020/10/012}

\bibitem[{{van Engelen} {et~al.}(2014){van Engelen}, {Bhattacharya}, {Sehgal},
  {Holder}, {Zahn}, \& {Nagai}}]{vanengelen14}
{van Engelen}, A., {Bhattacharya}, S., {Sehgal}, N., {et~al.} 2014, \apj, 786,
  13, \dodoi{10.1088/0004-637X/786/1/13}

\bibitem[{{van Engelen} {et~al.}(2012){van Engelen}, {Keisler}, {Zahn}, {Aird},
  {Benson}, {Bleem}, {Carlstrom}, {Chang}, {Cho}, {Crawford}, {Crites}, {de
  Haan}, {Dobbs}, {Dudley}, {George}, {Halverson}, {Holder}, {Holzapfel},
  {Hoover}, {Hou}, {Hrubes}, {Joy}, {Knox}, {Lee}, {Leitch}, {Lueker},
  {Luong-Van}, {McMahon}, {Mehl}, {Meyer}, {Millea}, {Mohr}, {Montroy},
  {Natoli}, {Padin}, {Plagge}, {Pryke}, {Reichardt}, {Ruhl}, {Sayre},
  {Schaffer}, {Shaw}, {Shirokoff}, {Spieler}, {Staniszewski}, {Stark}, {Story},
  {Vanderlinde}, {Vieira}, \& {Williamson}}]{vanengelen12}
{van Engelen}, A., {Keisler}, R., {Zahn}, O., {et~al.} 2012, \apj, 756, 142,
  \dodoi{10.1088/0004-637X/756/2/142}

\bibitem[{{Viero} {et~al.}(2013){Viero}, {Wang}, {Zemcov}, {Addison},
  {Amblard}, {Arumugam}, {Aussel}, {B{\'e}thermin}, {Bock}, {Boselli}, {Buat},
  {Burgarella}, {Casey}, {Clements}, {Conley}, {Conversi}, {Cooray}, {De
  Zotti}, {Dowell}, {Farrah}, {Franceschini}, {Glenn}, {Griffin},
  {Hatziminaoglou}, {Heinis}, {Ibar}, {Ivison}, {Lagache}, {Levenson},
  {Marchetti}, {Marsden}, {Nguyen}, {O'Halloran}, {Oliver}, {Omont}, {Page},
  {Papageorgiou}, {Pearson}, {P{\'e}rez-Fournon}, {Pohlen}, {Rigopoulou},
  {Roseboom}, {Rowan-Robinson}, {Schulz}, {Scott}, {Seymour}, {Shupe}, {Smith},
  {Symeonidis}, {Vaccari}, {Valtchanov}, {Vieira}, {Wardlow}, \&
  {Xu}}]{viero13}
{Viero}, M.~P., {Wang}, L., {Zemcov}, M., {et~al.} 2013, \apj, 772, 77,
  \dodoi{10.1088/0004-637X/772/1/77}

\bibitem[{{Wenzl} {et~al.}(in prep.)}]{dr6-cmass-eg}
{Wenzl}, L., {et~al.} in prep., To be submitted to MNRAS

\end{thebibliography}
\bibliographystyle{aasjournal}

\appendix

\section{Contribution to foreground bias from individual foreground components}

\begin{figure}
    \centering
    \includegraphics[width=0.45\textwidth]{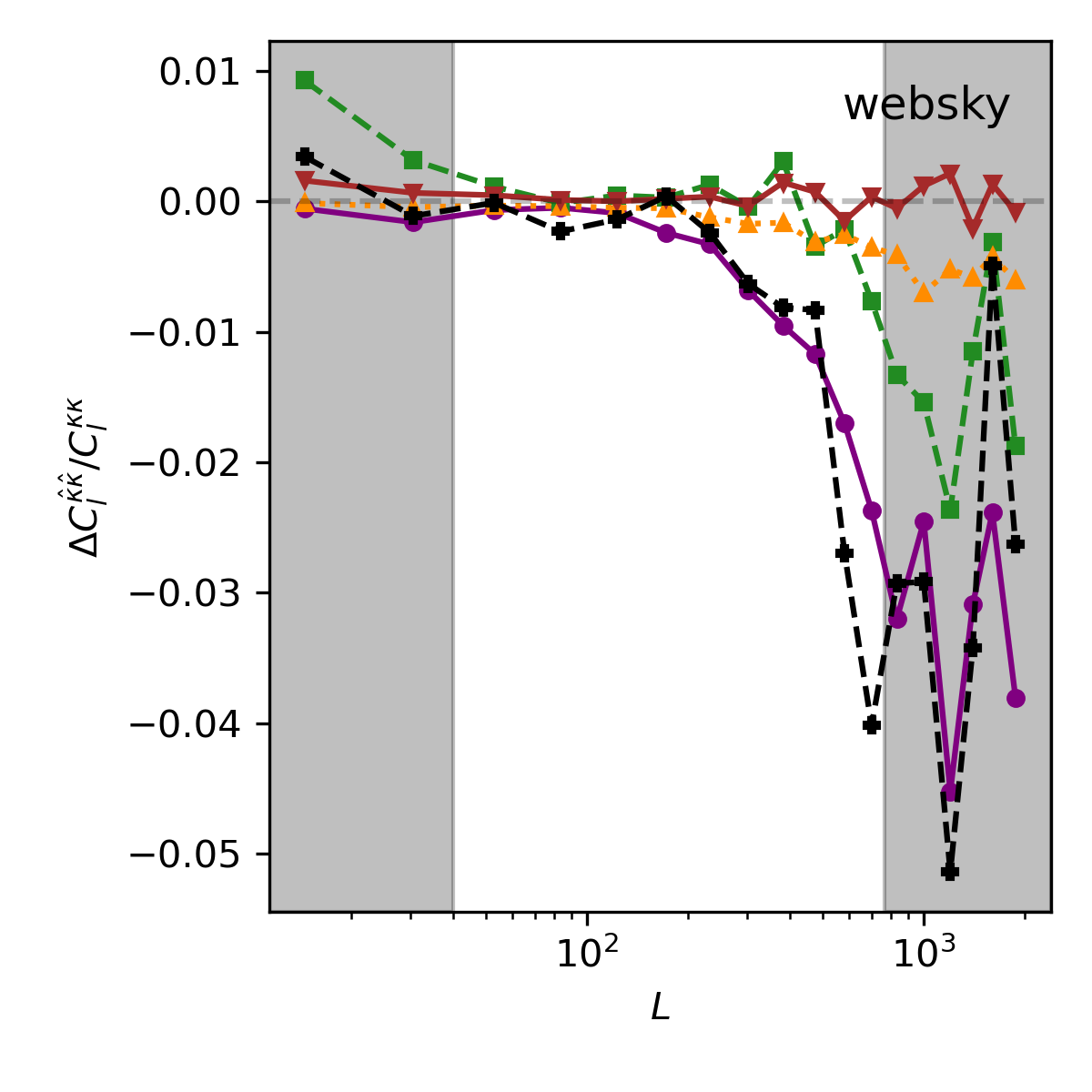}
    \includegraphics[width=0.45\textwidth]{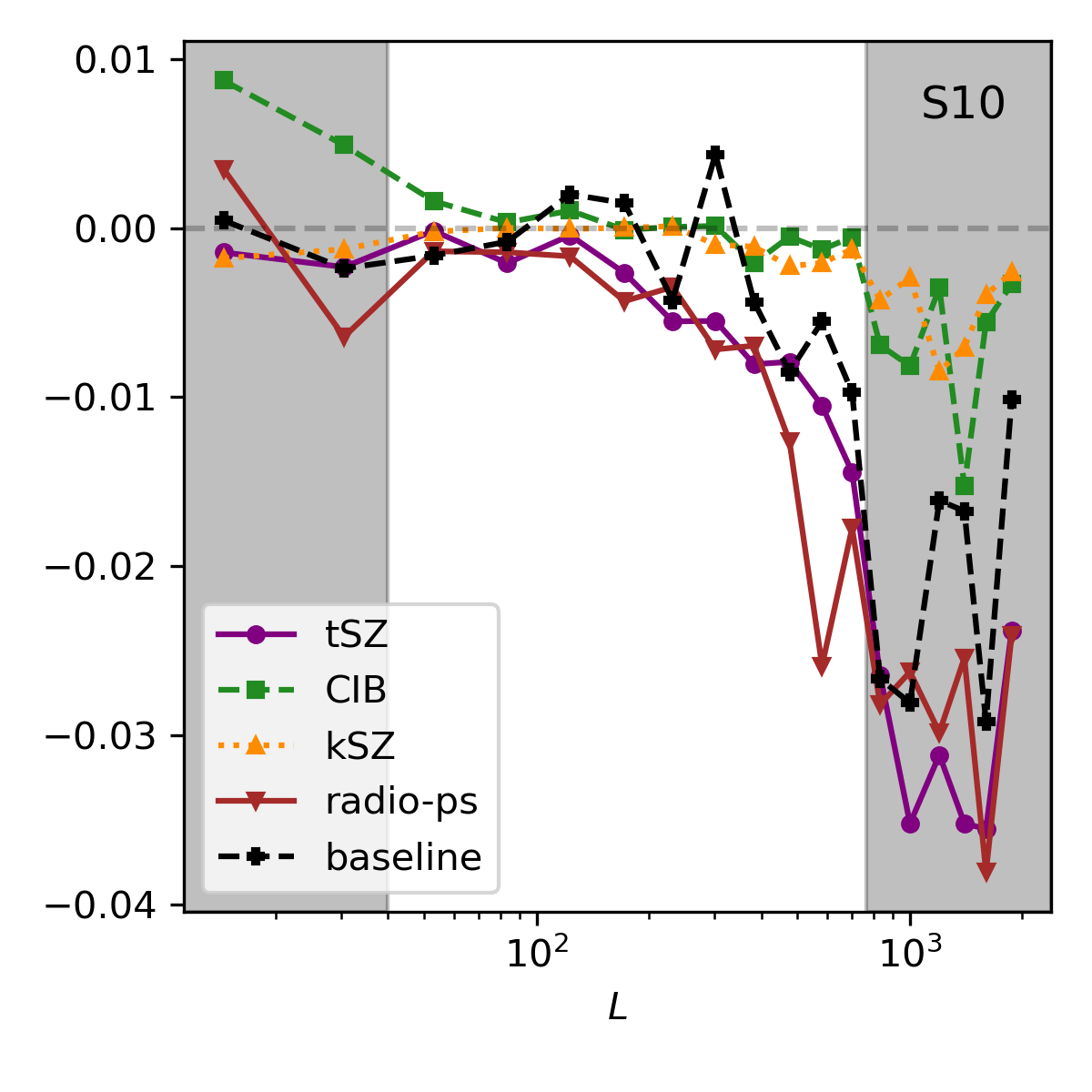}
    \caption{Contributions to the lensing power spectrum bias from individual foreground components: the thermal Sunyaev-Zeldovich effect (``tSZ''), the cosmic infrared background (``CIB''), the kinetic Sunyaev-Zeldovich effect (``kSZ''), and radio sources (``radio-ps''), as predicted from the \websky\ (left panel) and \sehgal\ (right panel) simulations. The total bias is labelled ``baseline'', and is not equal to the sum of the individual contributions, since there are also cross-terms present due to correlations on the sky between the different foregrounds (e.g. cosmic infrared background is correlated with tSZ).}
    \label{fig:ind_comp}
\end{figure}

We show in \fig{fig:ind_comp} the foreground biases for each individual extragalactic foreground component, for \websky\ (left panel) and \sehgal\ (right panel). These are estimated by re-running the simulation processing described in \sect{sec:actsim}, but in each case including only a single  foreground component in the maps. We also show the total bias as the black circular markers and dashed lines. Note that this is not simply a sum of the individual components since there are additional terms due to the correlation between the foreground components (e.g. CIB is quite correlated with tSZ). 

\section{Results with $1/f$ modulated noise}\label{app:noise_test}

Our simulation-based foreground bias  estimates depend on the effectiveness of point-source and cluster detection, which in turn depends on the properties of the noise added to the simulated foreground maps. Above we use a simple, local variance $1/\texttt{ivar}$  model for the map noise, where \texttt{ivar} is the inverse variance map esimtated for the DR6 coadd data). We  test here the inclusion of additional large-scale correlations by generating a simulated noise map as a Gaussian random field drawn from a power spectrum
$N_l = (1+ (l/l_{\mathrm{knee}})^\alpha)$, which is then multiplied by $\sqrt{1/\texttt{ivar}}$. On large scales ($l<l_{\mathrm{knee}}$), this introduces correlated noise that resembles that expected due to the atmosphere, while still achieving the correct pixel variance at small scales ($l>>l_{\mathrm{knee}}$). It is found to be a good fit to ACT data in \citet{naess20}, from where we take the parameter values  $l_{\mathrm{knee}} = (2000,3000)$ and $\alpha = (-3,-3)$ for the $(90, 150)$ GHz channels.

\begin{figure}
    \centering
    \begin{minipage}[t]{.48\textwidth}
    \centering
\includegraphics[width=0.95\textwidth]{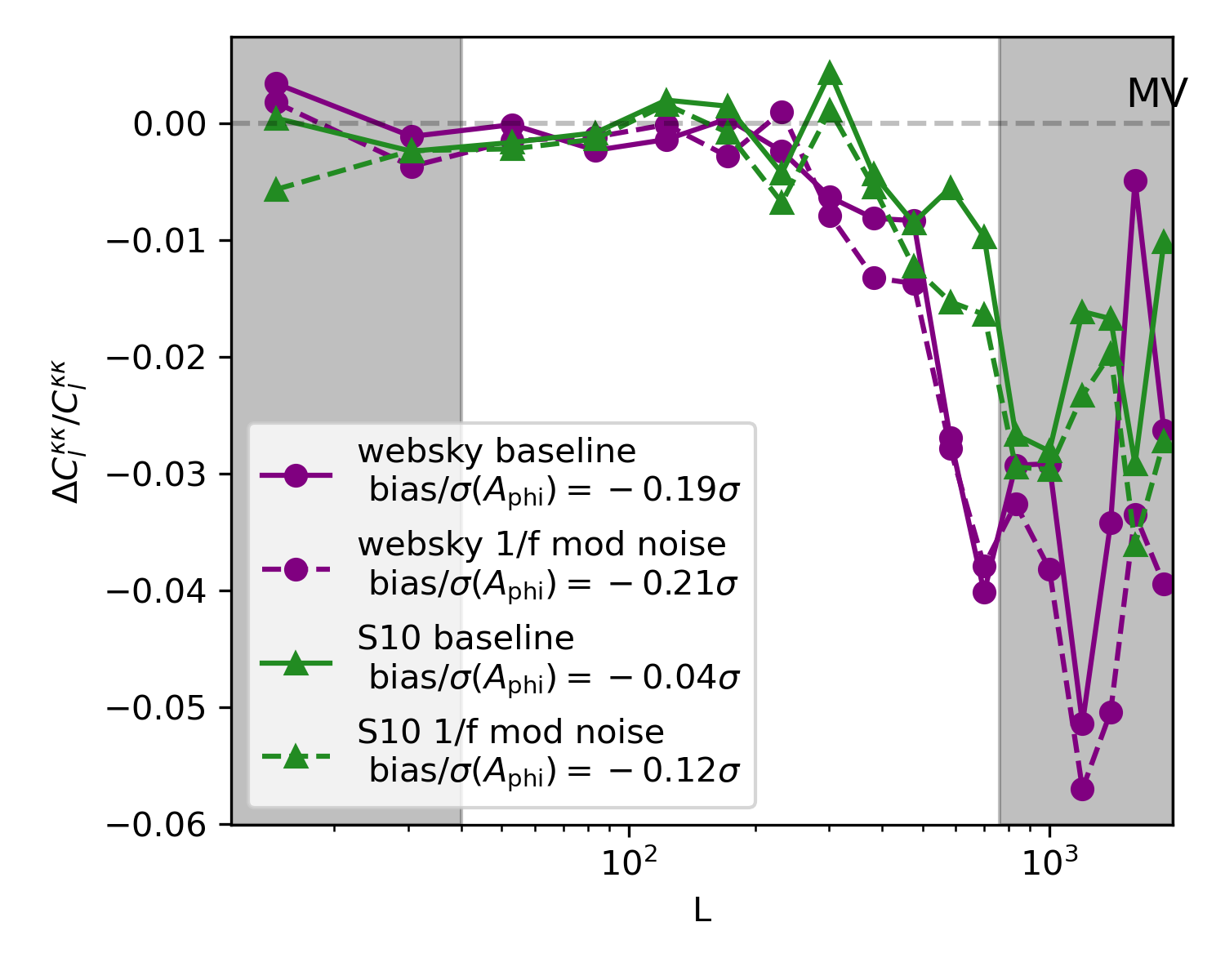}
    \caption{The dashed lines show predicted biases to the MV lensing reconstruction power spectrum, for the \websky\ and \sehgal\ simulations, when assuming the $1/f$ modulated noise model described in \app{app:noise_test} rather than our fiducial noise model (solid lines). In the legend we note the bias in inferred lensing amplitude as a fraction of the $1\sigma$ uncertainty for the ACT DR6 lensing analysis. The grey shaded regions indicate scales not used in the cosmological inference, as described in \citet{dr6-lensing-auto}.  Dashed grey lines indicate zero bias.}
    \label{fig:ttbias_1fnoise}
\end{minipage}%
\hfill
\begin{minipage}[t]{.48\linewidth}
    \centering
    \centering
    \includegraphics[width=0.95\textwidth]{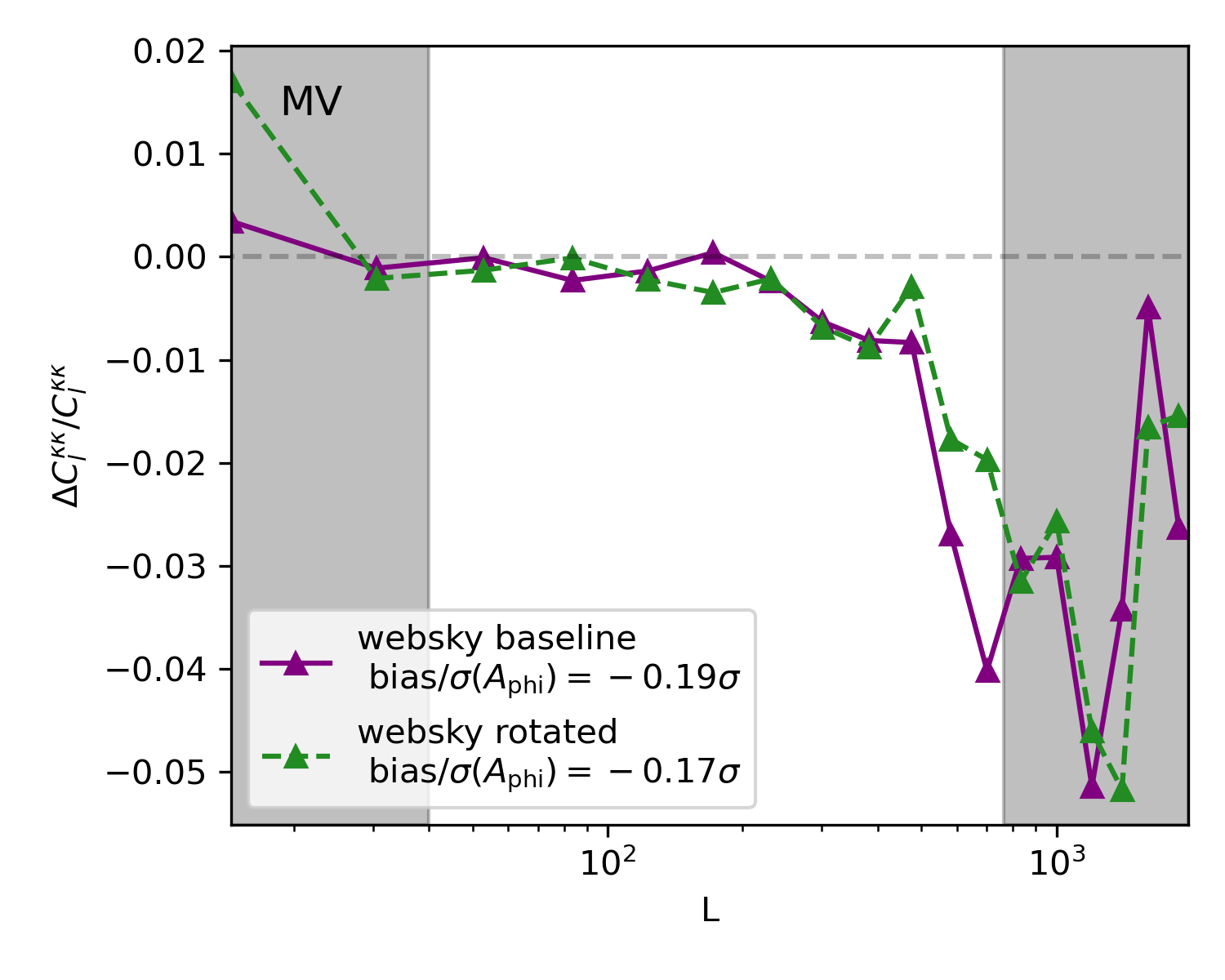}
    \caption{The dashed lines show predicted biases to the lensing reconstruction power spectrum for the rotated (as described in \sect{app:webskyrot}) \websky\ simulations, compared to the baseline case (solid line).  The grey shaded regions indicate scales not used in the cosmological inference, as described in \citet{dr6-lensing-auto}.}
    \label{fig:simbias_rot}
\end{minipage}
\end{figure}

\begin{figure}
    \centering
    \includegraphics[width=0.45\textwidth]{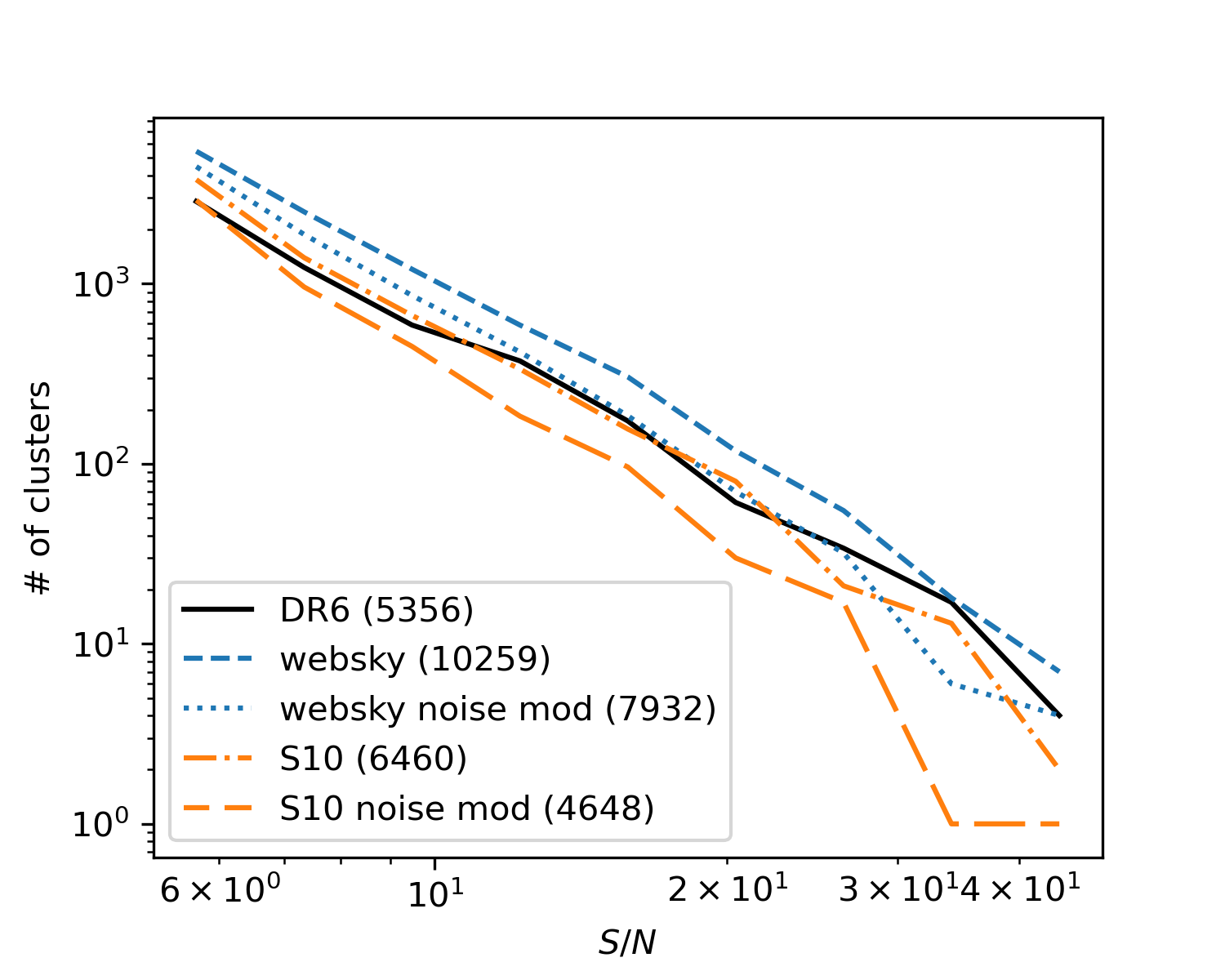}
    \includegraphics[width=0.45\textwidth]{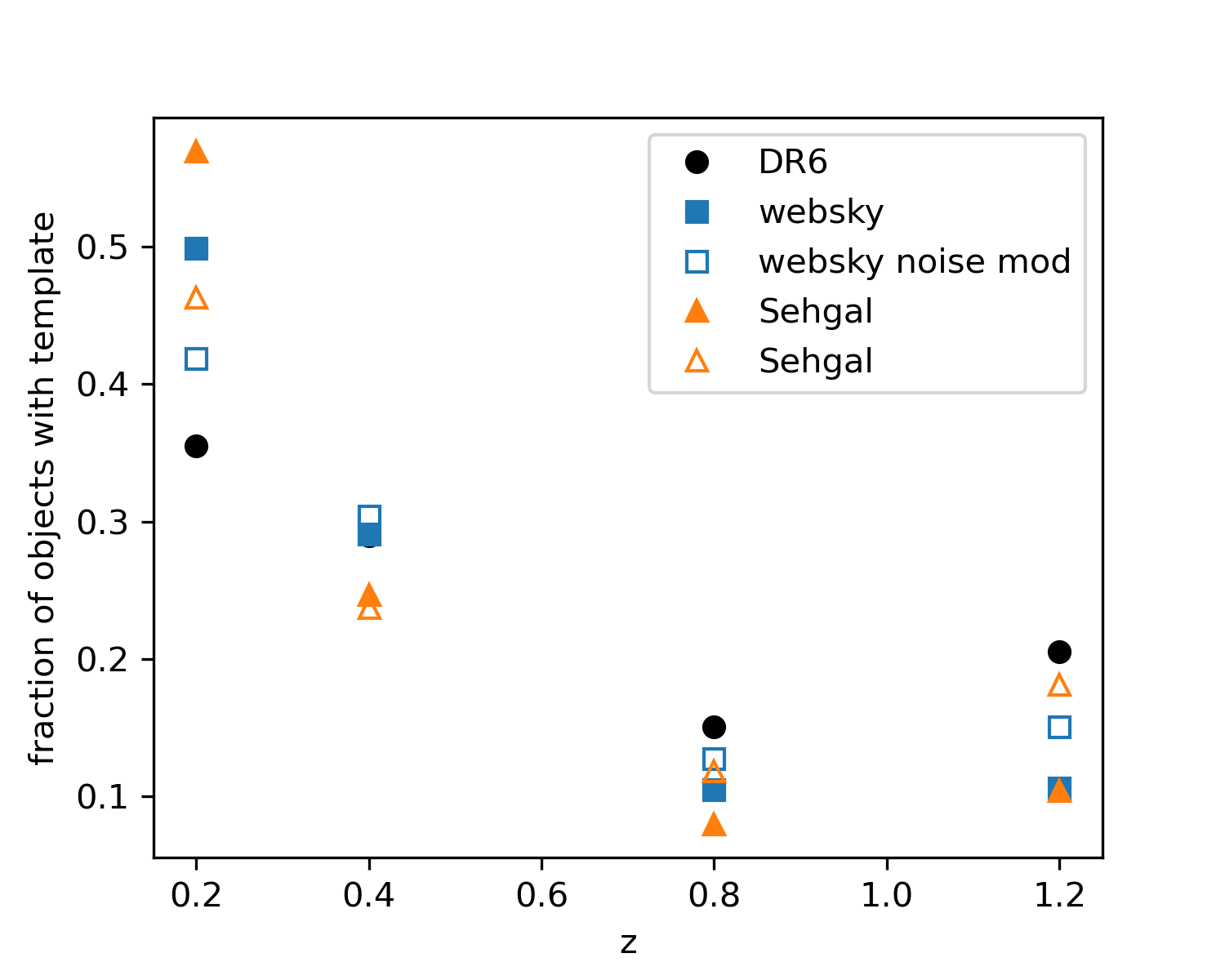}
    \caption{Left: The number of $S/N>5$ cluster candidates detected by \nemo\ as a function of $S/N$, for the DR6 data, and the \websky\ and \sehgal\ simulations. As well as our fiducial noise model, we show the number counts assuming the modulated $1/f$ noise model described in \app{app:noise_test}. The total number of $S/N>5$ cluster candidates for each case are noted in the legend. Right: Fraction of candidate clusters as a function of best-fit template redshift.}
    \label{fig:cluster_counts}
\end{figure}

As shown in \fig{fig:cluster_counts}, we do find that the results of the \textsc{nemo} cluster finding code are somewhat sensitive to this change in noise model, with fewer clusters found at low redshift (or at least, assigned low redshift best-fit templates), and somewhat fewer clusters detected in total. This is expected - the increased noise on large scales reduces the effectiveness in detecting the larger (in angular size) low-redshift clusters (e.g. see discussion in \citealt{Hilton_2021}). Nonentheless, there is little impact on the resulting foreground biases predicted for the lensing power spectrum, with negligible change to the bias in the infereed $\Alens$ (see \fig{fig:ttbias_1fnoise}).

\section{Uncertainty on  bias estimates due to finite simulation volume}\label{app:webskyrot}

In \sect{sec:simpredictions} we present predictions of the foreground biases to $\Clhatkappa{L}$ based on simulations. Unlike lensing estimation from a normal CMB map, these bias predictions are not affected by instrumental noise or noise on the CMB power spectrum. However, there does exist some uncertainty associated with the finite volume of simulation from which they are estimated. For both the \websky\ and \sehgal\ simulations one full-sky is available, and in order to ensure realistic noise properties for cluster and source finding, we further apply the ACT DR6 mask. We generate a close to independent 
(within the ACT mask) realization of the \websky\ simulation by rotating the simulation maps such that the sky area that enters the ACT DR6 mask does not overlap with that initially entering the ACT mask. A simple rotation by 90 degrees around the y-axis, implemented using pixell's\footnote{\url{https://github.com/simonsobs/pixell}} 
\texttt{rotate\_alm} function with angle arguments \texttt{(0., -np.pi/2, 0.)}, generates a map that has negligble overlap with the original area allowed by the mask. 

\begin{figure}
    \centering
    \includegraphics[width=0.5\textwidth]{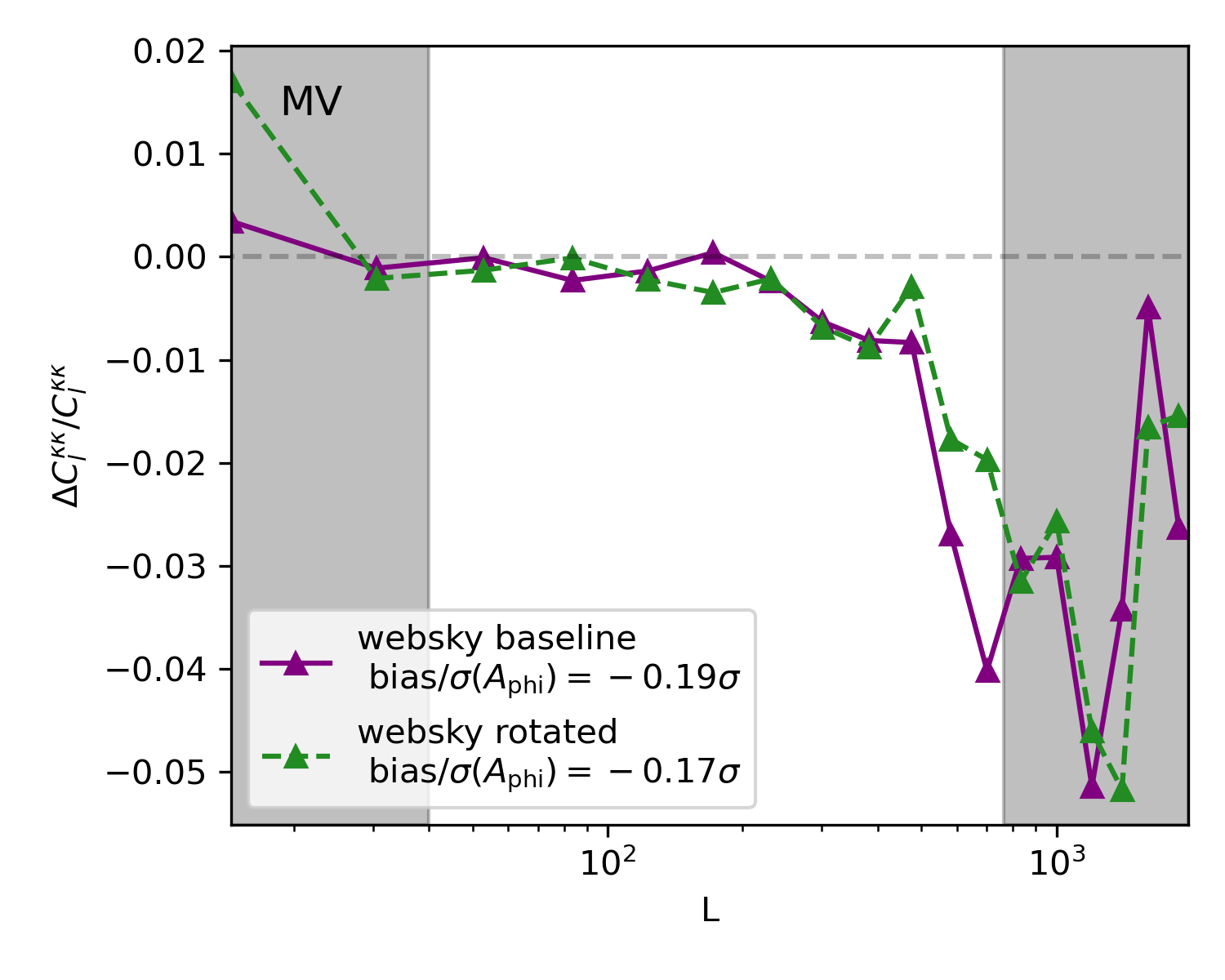}
    \caption{The dashed lines show predicted biases to the lensing reconstruction power spectrum for the rotated (as described in \sect{app:webskyrot}) \websky\ simulations, compared to the baseline case (solid line).  The grey shaded regions indicate scales not used in the cosmological inference, as described in \citet{dr6-lensing-auto}.}
    \label{fig:simbias_rot}
\end{figure}

The green dashed line in \fig{fig:simbias_rot} shows the foreground-induced bias for the MV case, with the bias for the rotated case well within requirements (an $\Alens$ bias of $-0.11\sigma$), and similar to the baseline (unrotated case), implying that cosmic variance is not a significant source of uncertainty in our foreground bias predictions.

\end{document}